\documentclass[a4paper,11pt]{article}
\pdfoutput=1 

\usepackage{jheppub} 

\usepackage[utf8]{inputenc}
\usepackage{graphicx} 
\usepackage{grffile}
\usepackage{subcaption}
\usepackage{dcolumn}  
\usepackage{makecell}
\usepackage{colordvi}
\usepackage{color}
\usepackage{epstopdf}
\usepackage{amssymb}
\usepackage{amsmath}
\usepackage{slashed}
\usepackage{enumerate}
\usepackage{url}
\usepackage{morefloats}

\usepackage{natbib}
\usepackage{accents}

\usepackage[colorlinks=true,
urlcolor=blue,
linkcolor=blue,
citecolor=blue,
linktocpage=true,
pdfproducer=medialab,
pdfa=true,
anchorcolor=blue]{hyperref}

\usepackage{tabularx,array}

\newcolumntype{C}{>{\centering\arraybackslash}X}
\usepackage{multirow}

\usepackage{lineno}

\usepackage{float}
\usepackage{xcolor}

\usepackage[title]{appendix}

\renewcommand{\arraystretch}{1.1}

\renewcommand{\thesection}{\arabic{section}}

\usepackage{changepage}
\usepackage{booktabs}
\newcommand{\tabitem}{~~\llap{\textbullet}~~}

\usepackage{orcidlink}

\RequirePackage{xspace}

\def\en         {\ensuremath{e^-}\xspace}   
\def\ep         {\ensuremath{e^+}\xspace}

\def\ee         {\ensuremath{e^-e^-}\xspace}

\def\Kbar  {\kern 0.2em\overline{\kern -0.2em K}{}\xspace}

\def\Kz    {\ensuremath{K^0}\xspace}
\def\Kzb   {\ensuremath{\Kbar^0}\xspace}
\def\KzKzb {\ensuremath{\Kz \kern -0.16em \Kzb}\xspace}
\def\Kp    {\ensuremath{K^+}\xspace}
\def\Km    {\ensuremath{K^-}\xspace}

\def\KpKm  {\ensuremath{\Kp \kern -0.16em \Km}\xspace}

\def\Dbar    {\kern 0.2em\overline{\kern -0.2em D}{}\xspace}

\def\Dz      {\ensuremath{D^0}\xspace}
\def\Dzb     {\ensuremath{\Dbar^0}\xspace}
\def\DzDzb   {\ensuremath{\Dz {\kern -0.16em \Dzb}}\xspace}
\def\Dp      {\ensuremath{D^+}\xspace}
\def\Dm      {\ensuremath{D^-}\xspace}

\def\DpDm    {\ensuremath{\Dp {\kern -0.16em \Dm}}\xspace}

\def\B       {\ensuremath{B}\xspace}
\def\Bbar    {\kern 0.18em\overline{\kern -0.18em B}{}\xspace}

\def\Bz      {\ensuremath{B^0}\xspace}
\def\Bzb     {\ensuremath{\Bbar^0}\xspace}
\def\BzBzb   {\ensuremath{\Bz {\kern -0.16em \Bzb}}\xspace}
\def\Bu      {\ensuremath{B^+}\xspace}
\def\Bub     {\ensuremath{B^-}\xspace}

\def\BpBm    {\ensuremath{\Bu {\kern -0.16em \Bub}}\xspace}

\mathchardef\Upsilon="7107
\def\Y#1S{\ensuremath{\Upsilon{(#1S)}}\xspace}

\def\FourS {\Y4S}

\mathchardef\Deltares="7101
\mathchardef\Xi="7104
\mathchardef\Lambda="7103
\mathchardef\Sigma="7106
\mathchardef\Omega="710A

\def\Deltabar{\kern 0.25em\overline{\kern -0.25em \Deltares}{}\xspace}
\def\Lbar{\kern 0.2em\overline{\kern -0.2em\Lambda\kern 0.05em}\kern-0.05em{}\xspace}
\def\Sigbar{\kern 0.2em\overline{\kern -0.2em \Sigma}{}\xspace}
\def\Xibar{\kern 0.2em\overline{\kern -0.2em \Xi}{}\xspace}
\def\Obar{\kern 0.2em\overline{\kern -0.2em \Omega}{}\xspace}
\def\Nbar{\kern 0.2em\overline{\kern -0.2em N}{}\xspace}
\def\Xb{\kern 0.2em\overline{\kern -0.2em X}{}\xspace}

\newcommand{\tev}{\ensuremath{\mathrm{\,Te\kern -0.1em V}}\xspace}
\newcommand{\gev}{\ensuremath{\mathrm{\,Ge\kern -0.1em V}}\xspace}
\newcommand{\mev}{\ensuremath{\mathrm{\,Me\kern -0.1em V}}\xspace}
\newcommand{\kev}{\ensuremath{\mathrm{\,ke\kern -0.1em V}}\xspace}
\newcommand{\ev}{\ensuremath{\mathrm{\,e\kern -0.1em V}}\xspace}
\newcommand{\gevc}{\ensuremath{{\mathrm{\,Ge\kern -0.1em V\!/}c}}\xspace}
\newcommand{\mevc}{\ensuremath{{\mathrm{\,Me\kern -0.1em V\!/}c}}\xspace}
\newcommand{\gevcc}{\ensuremath{{\mathrm{\,Ge\kern -0.1em V\!/}c^2}}\xspace}
\newcommand{\mevcc}{\ensuremath{{\mathrm{\,Me\kern -0.1em V\!/}c^2}}\xspace}

\def\invfb   {\ensuremath{\mbox{\,fb}^{-1}}\xspace}

\def\mus  {\ensuremath{\rm \,\mus}\xspace}

\def\mus        {\ensuremath{\,\mu{\rm s}}\xspace}    

\def\to                 {\ensuremath{\rightarrow}\xspace}

\def\gsim{{~\raise.15em\hbox{$>$}\kern-.85em
          \lower.35em\hbox{$\sim$}~}\xspace}
\def\lsim{{~\raise.15em\hbox{$<$}\kern-.85em
          \lower.35em\hbox{$\sim$}~}\xspace}

\def\onres{on-resonance\xspace}
\def\conti{off-resonance\xspace}
\def\sqrts{\ensuremath{\sqrt{s}}\xspace}

\def\ee{\ensuremath{e^+e^-}\xspace}
\def\qq{\ensuremath{q\bar{{q}}}\xspace}

\def\ntrkoff{\ensuremath{{\rm N}_{\rm trk}^{\rm rec}}\xspace}
\def\bzz{\ensuremath{B(0,0)}\xspace}
\def\deta{\ensuremath{\Delta\eta}\xspace}
\def\dphi{\ensuremath{\Delta\phi}\xspace}
\def\ydphi{\ensuremath{Y(\Delta \phi)}\xspace}

\begin{document}

\rightline{\vbox{ \hbox{Belle Preprint 		2022-11}
                  \hbox{KEK Preprint 		2022-10}
}}

\title{Two-particle angular correlations in \boldmath $e^+ e^-$ collisions to hadronic final states in two reference coordinates at Belle}

\collaboration{The Belle Collaboration}
  \author{Y.-C.~Chen\,\orcidlink{0000-0002-9038-5324},} 
  \author{Y.-J.~Lee\,\orcidlink{0000-0003-2593-7767},} 
  \author{P.~Chang\,\orcidlink{0000-0003-4064-388X},} 
  \author{I.~Adachi\,\orcidlink{0000-0003-2287-0173},} 
  \author{H.~Aihara\,\orcidlink{0000-0002-1907-5964},} 
  \author{S.~Al~Said\,\orcidlink{0000-0002-4895-3869},} 
  \author{D.~M.~Asner\,\orcidlink{0000-0002-1586-5790},} 
  \author{H.~Atmacan\,\orcidlink{0000-0003-2435-501X},} 
  \author{V.~Aulchenko\,\orcidlink{0000-0002-5394-4406},} 
  \author{T.~Aushev\,\orcidlink{0000-0002-6347-7055},} 
% ...

\date{\today}

\abstract{
We present the analysis of two-particle angular correlations using coordinate systems defined with the conventional beam axis and the event thrust axis. We propose the latter as a good representation for the correlation structure interpretation in the \ee collision system. The  \ee collisions to hadronic final states at center-of-mass energies of $\sqrt{s} = 10.52$ GeV and 10.58 GeV are recorded by the Belle detector at KEKB.
In this paper, results on the first dataset are supplementary to the previous Belle publication~\cite{Belle:2022fvl}. At the same time, the latter is the first two-particle correlation measurement at collision energy on the $\Upsilon(4S)$ resonance and is sensitive to its decay products.
Measurements are reported as a function of the charged-particle multiplicity.
Finally, a qualitative understanding of the correlation structure is discussed using a combination of Monte Carlo simulations and experimental data.
}
\maketitle
\flushbottom


\section{Introduction}
\label{sec:Introduction}
In nucleon-nucleon and heavy-ion collision experiments, two-particle angular correlations~\cite{STAR:2005ryu,STAR:2009ngv,PHOBOS:2009sau,Chatrchyan:2012wg,Aamodt:2011by,Adam:2019woz} are extracted for the study of the Quark-Gluon Plasma (QGP)~\cite{Busza:2018rrf}, color reconnection~\cite{OrtizVelasquez:2013ofg}, color rope interaction~\cite{Bierlich:2020naj} and the search for initial-state correlation effects such as the Color Glass Condensate~\cite{Dumitru:2010iy}.
In these measurements, a long-range angular correlation, the ridge-like structure~\cite{STAR:2009ngv,PHOBOS:2009sau}, has been observed in various collision systems and at various collision energies. Since the beginning of the LHC operation, a ridge structure in the near-side ($\Delta\phi\approx 0$ between particles) has been observed also in high-multiplicity proton-proton collisions by the CMS collaboration~\cite{Khachatryan:2010gv} and confirmed by experiments at LHC and RHIC using smaller collision systems such as proton-ion~\cite{CMS:2012qk,Aaij:2015qcq} collisions. The studies have later been extended to the away-side ($\Delta\phi\approx \pi$) in proton-proton~\cite{Aad:2015gqa}, proton-ion~\cite{Abelev:2012ola,ATLAS:2012cix} and deuteron-ion~\cite{PHENIX:2013ktj} collisions.
In heavy-ion collisions, the ridge structure is associated with the hydrodynamic expansion of a strongly interacting and expanding system~\cite{Ollitrault:1992bk} produced from collisions at different impact parameters and the fluctuating initial state~\cite{Alver:2010gr}. However, the physical origin of the ridge structure in small systems is still under debate~\cite{Dusling:2013qoz,Bozek:2011if,He:2015hfa,Nagle:2018nvi}.
Recently, there has been growing interest in measuring two-particle correlations in even smaller collision systems of $ep$~\cite{ZEUS:2019jya} and \ee~\cite{Nagle:2017sjv,Badea:2019vey,Belle:2022fvl}, as they serve as counterparts complementary to the results in large collision systems and can be used to find the minimal conditions for collective behavior~\cite{Nagle:2017sjv}. The use of an electron beam removes complications such as multiple parton interactions and initial-state correlations. Those measurements have provided additional insights on the ridge signal~\cite{Altinoluk:2020wpf,Bierlich:2020naj,Castorina:2020iia,Agostini:2021xca}.

In this work, two-particle correlations calculated using a coordinate system with the beam pipe direction assigned as the $z$ axis are reported in full detail.
However, as pointed out in the study of ALEPH \ee collisions via hadronic $Z$ decay data~\cite{Badea:2019vey}, this conventional coordinate system might not be the optimal configuration in the \ee collision system.
In the \ee system, back-to-back dijets fragmenting from the $q\bar{q}$ pair production dominate the event topology, and the directions of such fragmentation are not fixed but emitted with a $1+\cos^2\theta$ angular distribution relative to the beam pipe. 
Thus, pair-wise correlations of the global dijet topology contribute differently to the two-particle angular correlation functions from one event to another, depending on the orientation of the dijet in each event.
Such smearing effects can degrade the detection sensitivity of the anomalous particle collectivity originating from the possibly thermalized medium, which is, if any, intuitively assumed to take place between the two out-going quarks.
In the \ee collision system, without the complexity of parton distribution functions and large underlying events, one can mitigate the smearing effect to allow for an understanding of finer correlation structures by defining the angular correlations relative to the event thrust axis, which is a well-defined surrogate of the hard process quark-pair direction using final-state particles.
On the other hand, it is hard to conclude the overall direction of the parton-level interaction with a single axis in hadron-hadron collision.

In this paper, we present a detailed description of the analysis methodology and results for the thrust-axis two-particle correlation function measurement.
Measurement with the thrust axis analysis provides a different view from the conventional beam axis analysis. For the \ee annihilation process, the thrust axis analysis is less descriptive of the leading dijet correlation while more sensitive to higher-order soft emissions.

We report on the measurement of two-particle angular correlation functions in high-multiplicity $e^+e^-$ annihilation events at $\sqrt{s}= 10.52$~GeV and 10.58~GeV data taken by the Belle experiment.
The center-of-mass energy of the second dataset is set on the $\Upsilon(4S)$ resonance, with about one-fourth the events decaying from $\Upsilon(4S)$ states into $B\bar{B}$ mesons. The two-particle correlation of the latter dataset is studied for the first time. 
This measurement is sensitive to anisotropic correlations in the azimuthal angle due to the resonance decays in the dataset, which we further discuss aided by Monte Carlo simulations.
The other dataset is 60~MeV below the resonance, and mostly composed of quark pairs fragmentating into hadrons.
Those are tests of the existence of ridge-like signals without the complications from initial states.
This study could also provide significant new input to the phenomenological fragmentation models at the low-energy regime of the Belle data.
Taking advantage of the high statistics of the data, an extreme charged-particle multiplicity reach of around 14 particles may be achieved, corresponding to 0.04\% of all hadronic \ee annihilation events and being about twice the average multiplicity.
In addition, we use simulations to better understand the correlation structure and a comparison of the expectations from different event generators, offering insights to the fragmentation models and a deeper look into the explanation of two-particle correlation.

The results are partially reported in ref.~\cite{Belle:2022fvl}; this paper puts more emphasis on introducing the thrust-axis two-particle correlations analysis methodology and understanding the measured results by means of simulations.
The presentation is organized as follows: The data sample and corresponding event thrust distributions are reported in section~\ref{sec:Sample}. Descriptions of the thrust-axis two-particle correlation function and the corrections applied are given in sections~\ref{sec:TwoParticleCorrelationFunction} and~\ref{sec:Corrections}. Two-particle correlations and the projected azimuthal differential associated yields are shown in section~\ref{sec:Rst}, as well as an estimate of the ridge-signal upper limits. 
With Monte Carlo (MC) event generators, we compare and study correlation functions in order to further understand the \ee collision at the low-energy regime in section~\ref{sec:GeneratorStudy}.

\section{Experimental setup, data sample and the event thrust}
\label{sec:Sample}

We perform two-particle correlation measurements with Belle \ee collision data. 
Two datasets, one at $\sqrt{s}=10.52$~GeV and the other at 10.58~GeV are analyzed, corresponding to about 89.5 \invfb and 332.2 \invfb of integrated luminosity, respectively. 
The $\sqrt{s}=10.58$ GeV collision data is targeted at the \FourS resonance and referred to as the ``on-resonance data'', while the other is called ``off-resonance data''.
Partial results for the \conti dataset are released~\cite{Belle:2022fvl} with the intention of providing a complementary understanding of two-particle correlations measured in hadron collisions with results from  high-energy \ee collision.
The full measurement and the analysis details are documented in this paper. 

The Belle experiment is configured with asymmetric 8~GeV electron and 3.5~GeV positron beams, customized to study the decay-time-dependent $C\hspace{-0.04em}P$ violation in $B$ decays. Data are collected with the Belle detector~\cite{Abashian:2000cg,Belle:2012iwr}, which is a large-solid-angle magnetic spectrometer consisting of a silicon vertex detector (SVD), a 50-layer central drift chamber (CDC), an array of aerogel threshold Cherenkov counters (ACC), a barrel-like arrangement of time-of-flight scintillation counters (TOF), and an electromagnetic calorimeter (ECL) comprising CsI(Tl) crystals, all located inside a superconducting solenoid coil that provides a 1.5~T magnetic field.  An iron flux-return surrounding the coil is instrumented to detect $K_L^0$ mesons and muons (KLM).

The \textsc{evtgen}~\cite{Ryd:2005zz}- and \textsc{pythia6}~\cite{2001CoPhC.135..238S}- based Belle MC sample is used as the simulation of $e^+e^-$ annihilation events, including hadronic $q\bar{q}$ ($q=u,d,s$ and $c$) fragmentation, \FourS decays, radiative Bhabha events, low multiplicity $e^{+}e^{-} \to l^{+}l^{-}$ ($l = e$ or $\mu$) and two-photon processes.
This MC dataset is used to study reconstruction  inefficiencies and thereby deriving efficiency correction factors for the data.
In the following, results with the Belle MC sample are also provided accompanying data results.

\begin{figure}[ht]
\centering
    \begin{subfigure}[b]{0.4\textwidth}
        \includegraphics[width=\textwidth,angle=0]{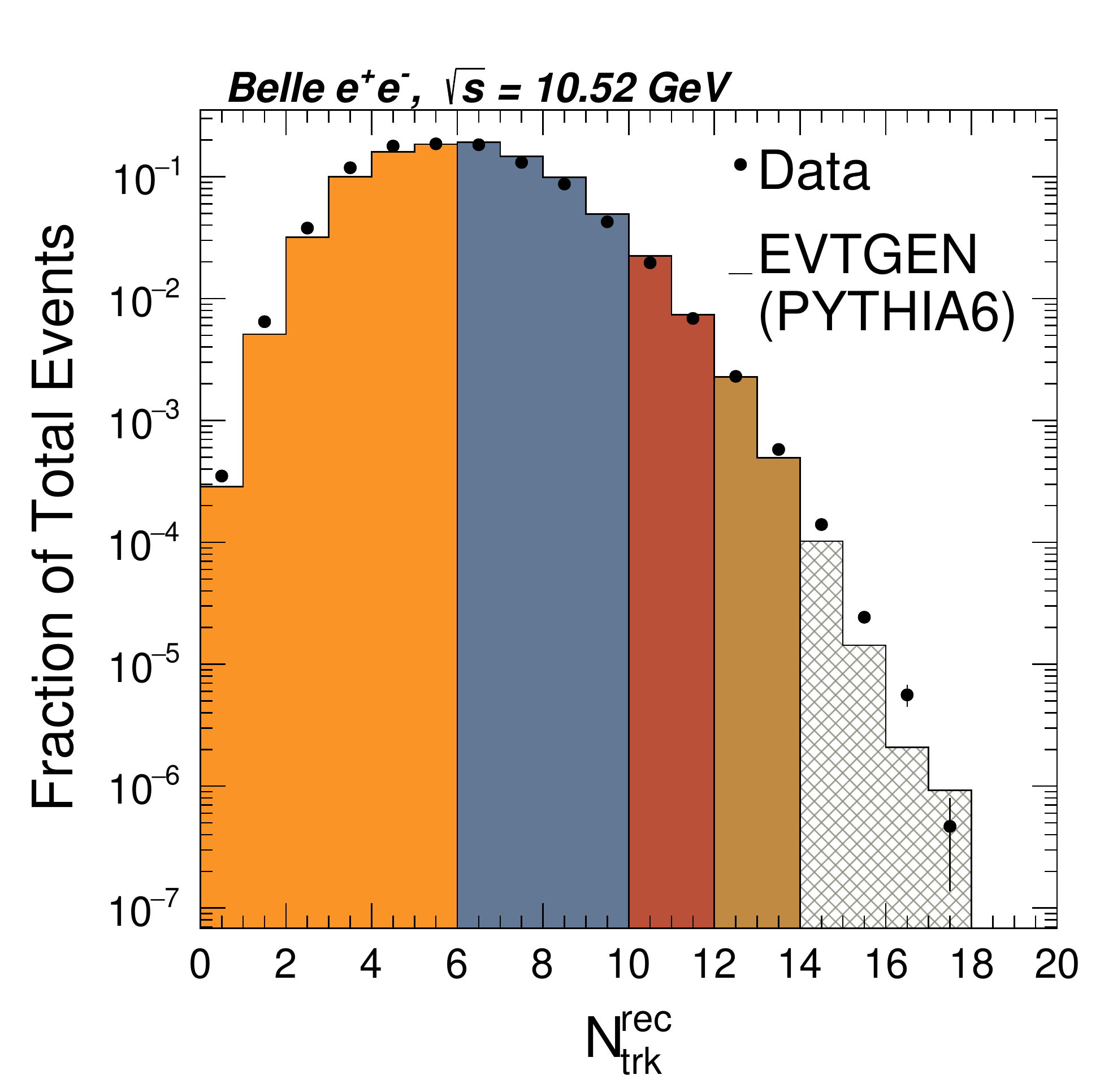}
        \caption{off-resonance}
    \end{subfigure}
    \begin{subfigure}[b]{0.4\textwidth}
        \includegraphics[width=\textwidth,angle=0]{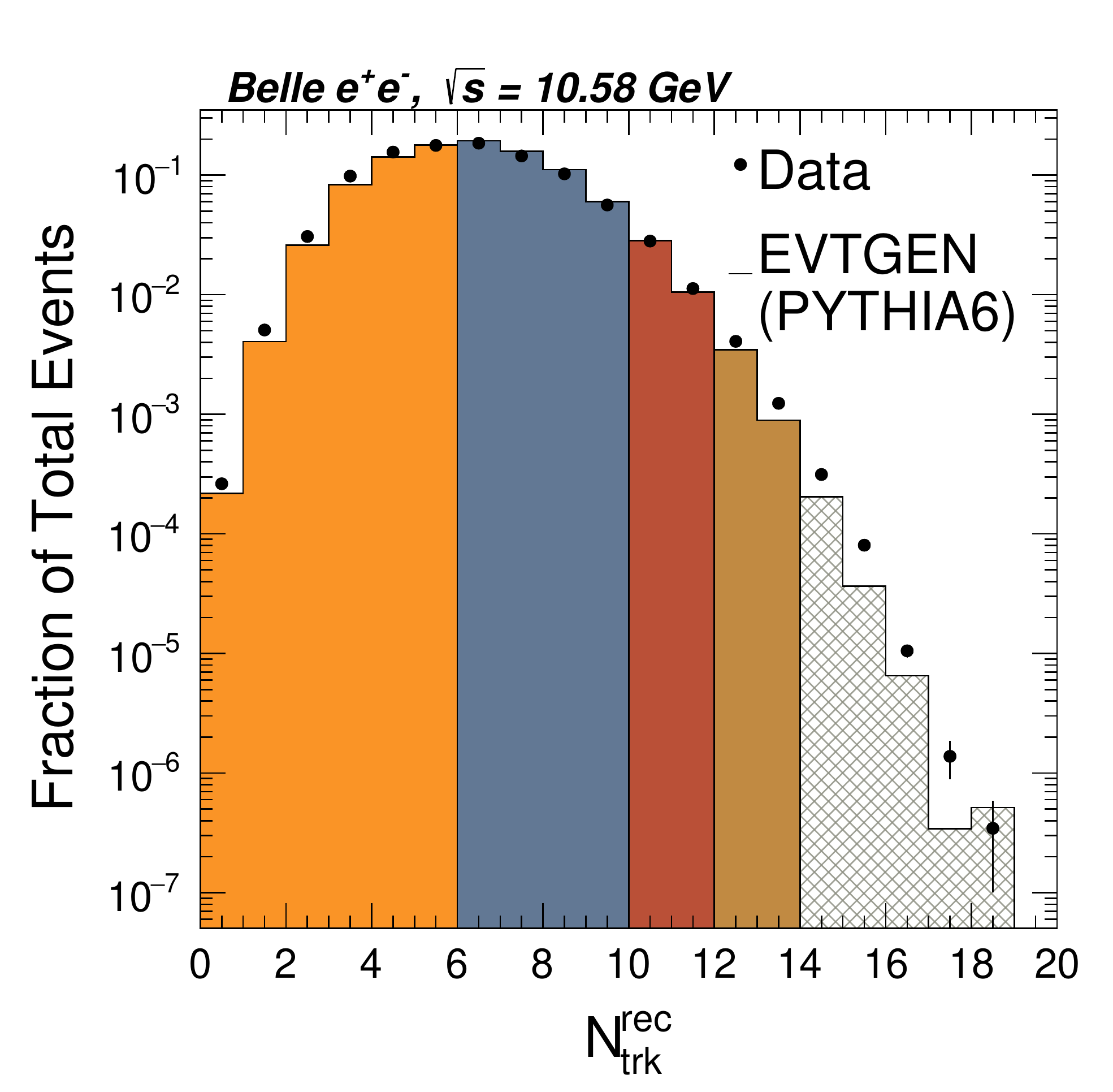}
        \caption{on-resonance}
    \end{subfigure}
\caption{The reconstructed-track multiplicity (${\rm N}_{\rm trk}^{\rm rec}$) distributions for (a) off-resonance and (b) on-resonance samples. The corresponding integrated luminosity used for plotting is about 1.3 $\rm fb^{-1}$ for each. 
Results of the data (black dots) and the \textsc{pythia6}-based Belle MC (histograms) are overlaid.
The colored bins in gold, blue, red, brown and gray are the events with ${\rm N}_{\rm trk}^{\rm rec}$ in $[0,6)$, $[6,10)$, $[10,12)$, $[12,14)$ and $[14,\infty)$ intervals.}
\label{fig:NtrkOffline}
\end{figure}

Primary charged particles (table~\ref{tab:SelectionSummary}) in hadronic events~\cite{Belle:2001jqo} are used for the two-particle correlation study.
We require those particles to be within the detector's angular coverage, and apply a loose requirement on the track's distance with respect to the interaction point to ensure the selected tracks are from collisions. We focus on high-momentum tracks with the transverse momentum $p_{\rm T}>0.2$~GeV$/c$. Reconstruction effects such as duplicate low-$p_{\rm T}$ tracks and electron-positron pairs from photon conversions are also studied, with specific removal treatments listed in table~\ref{tab:SelectionSummary}.
After all selections, the subsequent reconstructed-track multiplicity (\ntrkoff) is used for studying the correlation function's multiplicity dependence.
Considering the computational cost, events with $\ntrkoff < 12$ are analyzed with 11.5 \invfb and 11.7 \invfb partial samples for the \conti and \onres dataset, respectively, whereas the full datasets are used for analyzing the rarer high-multiplicity events.
Figure~\ref{fig:NtrkOffline} shows the reconstructed-track multiplicity for the \conti and \onres datasets. 
The multiplicity classes used in this study, their corresponding fraction of data, and the mapping of average reconstructed  multiplicities $\langle {\rm N}_{\rm trk}^{\rm rec}\rangle$ to average multiplicities after efficiency correction $\langle {\rm N}_{\rm trk}^{\rm corr}\rangle$ are listed in table~\ref{tab:NtrkCorr}. The efficiency correction procedure is later introduced in section~\ref{sec:EfficiencyCorrection}.

\begin{table}[ht]
\caption{Summary table for particle selections. Track selections have been elucidated in ref.~\cite{Belle:2022fvl}. Neutral particles are selected mainly for the event thrust calculation.}
\begin{center}
\begin{tabularx}{\textwidth}{l|l}
\hline\hline
\multicolumn{2}{l}{Charged particles}  \\
\hline
Primary tracks          & \tabitem [reconstructed] decay promptly or from the long-lifetime \\
                        & particle whose decay vertex from the interaction point is \\
                        & less than 1 cm in the transverse plane\\
                        & \qquad Long-lifetime particle candidates:\\
                        & \qquad \tabitem $K_S^0$ ($0.480$-$0.516$~GeV$/c^2$)\\
                        & \qquad \tabitem $\Lambda^0/\bar{\Lambda}^0$ ($1.111$-$1.121$~GeV$/c^2$)\\
                        & \tabitem [MC-truth] decay promptly or from particles with proper \\
                        & lifetime $\tau \leq 1$ cm/$c$ \\
Acceptance              & particles' polar angle (defined with respect to the opposite \\
                        & direction of the $e^+$ beam): $17^{\circ} \le \theta \le 150^{\circ}$ \\
High quality tracks     & $p_{\rm T} \ge 0.2$ GeV$/c$ \\
Impact parameter        & (tracks' radial and $z$- distance from the interaction point)\\ 
                        & $|\Delta r|<2$~cm, $|\Delta z|<5$~cm \\
Duplicate track removal & veto the softer track of a low-momentum pair ($p_{\rm T} < 0.4$ GeV$/c$) \\
                        & travelling with a small opening angle $\delta$: \\
                        & \hspace{0.3cm} 1. same-sign charges with $\cos\delta > 0.95$ \\
                        & \hspace{0.3cm} 2. opposite-sign charges with $\cos\delta < -0.95$ \\
Photon conversion veto  & veto track pairs which can form common vertices ($V^0$ objects) \\
                        & with \\
                        &\hspace{0.3cm} 1. $z$ distance between two tracks $<10$ cm  \\
                        &\hspace{0.3cm} 2. reconstructed $V^0$'s mass $< 0.25$ GeV$/c^2$  \\
                        &\hspace{0.3cm} 3. decay-vertex radius $> 1.5$ cm\\
\hline
\multicolumn{2}{l}{Neutral particles (for thrust calculation)}  \\
\hline
Cluster selection       & No association from tracks' extrapolations \\
Acceptance              & particles' polar angle (defined with respect to the opposite \\
                        & direction of the $e^+$ beam): $17^{\circ} \le \theta \le 150^{\circ}$ \\
Energy requirement      & Forward endcap: \hspace{0.10cm} $E \ge 0.10$ GeV \\
                        & Backward endcap:                $E \ge 0.15$ GeV \\
                        & Barrel:         \hspace{1.74cm} $E \ge 0.05$ GeV \\
\hline\hline
\end{tabularx} 
\label{tab:SelectionSummary}
\end{center}
\end{table} 

\clearpage

\begin{table}[ht]
\caption{Average multiplicities and corrected multiplicities with high-quality tracks for different ${\rm N}_{\rm trk}^{\rm rec}$ intervals for the \conti and \onres datasets. These tracks pass requirements in table~\ref{tab:SelectionSummary}: $p_{T} \ge 0.2$~GeV/$c$ and $17^\circ \le \theta \le 150^\circ$, hence the high efficiency.}
\begin{center}
\footnotesize
\begin{tabularx}{\textwidth}{
    >{\centering}p{0.09\textwidth} | 
    >{\centering}p{0.12\textwidth}
    >{\centering}p{0.12\textwidth}
    >{\centering}p{0.12\textwidth} |
    >{\centering}p{0.12\textwidth}
    >{\centering}p{0.15\textwidth}
    X
}
\hline\hline
\multirow{2}{*}{\parbox{0.09\textwidth}{${\rm N}_{\rm trk}^{\rm rec}$ interval}} &
\multicolumn{3}{c|}{Off-resonance} & 
\multicolumn{3}{c}{On-resonance} \\
\cline{2-7}
&  Fraction (\%) & $\left<{\rm N}_{\rm trk}^{\rm rec}\right>$ & $\left<{\rm N}_{\rm trk}^{\rm corr}\right>$ &   Fraction (\%) & $\left<{\rm N}_{\rm trk}^{\rm rec}\right>$ & {\footnotesize $\left<{\rm N}_{\rm trk}^{\rm corr}\right>$} \\
\hline 
$[6,10)$        & 44.33 &  6.98 &  7.05 & 48.80 &  7.06 &  {\footnotesize 7.18} \\
$[10,12)$       &  2.65 & 10.26 & 10.12 &  3.93 & 10.29 & {\footnotesize 10.20} \\
$[12,14)$       &  0.29 & 12.20 & 11.90 &  0.53 & 12.23 & {\footnotesize 11.99} \\
$[14,\infty)$   &  0.02 & 14.22 & 14.24 &  0.04 & 14.26 & {\footnotesize 13.93} \\
\hline\hline
\end{tabularx}
\label{tab:NtrkCorr}
\end{center}
\end{table}

\begin{figure}[ht]
    \begin{subfigure}[b]{\textwidth}
	\centering
	\includegraphics[width=0.8\textwidth]{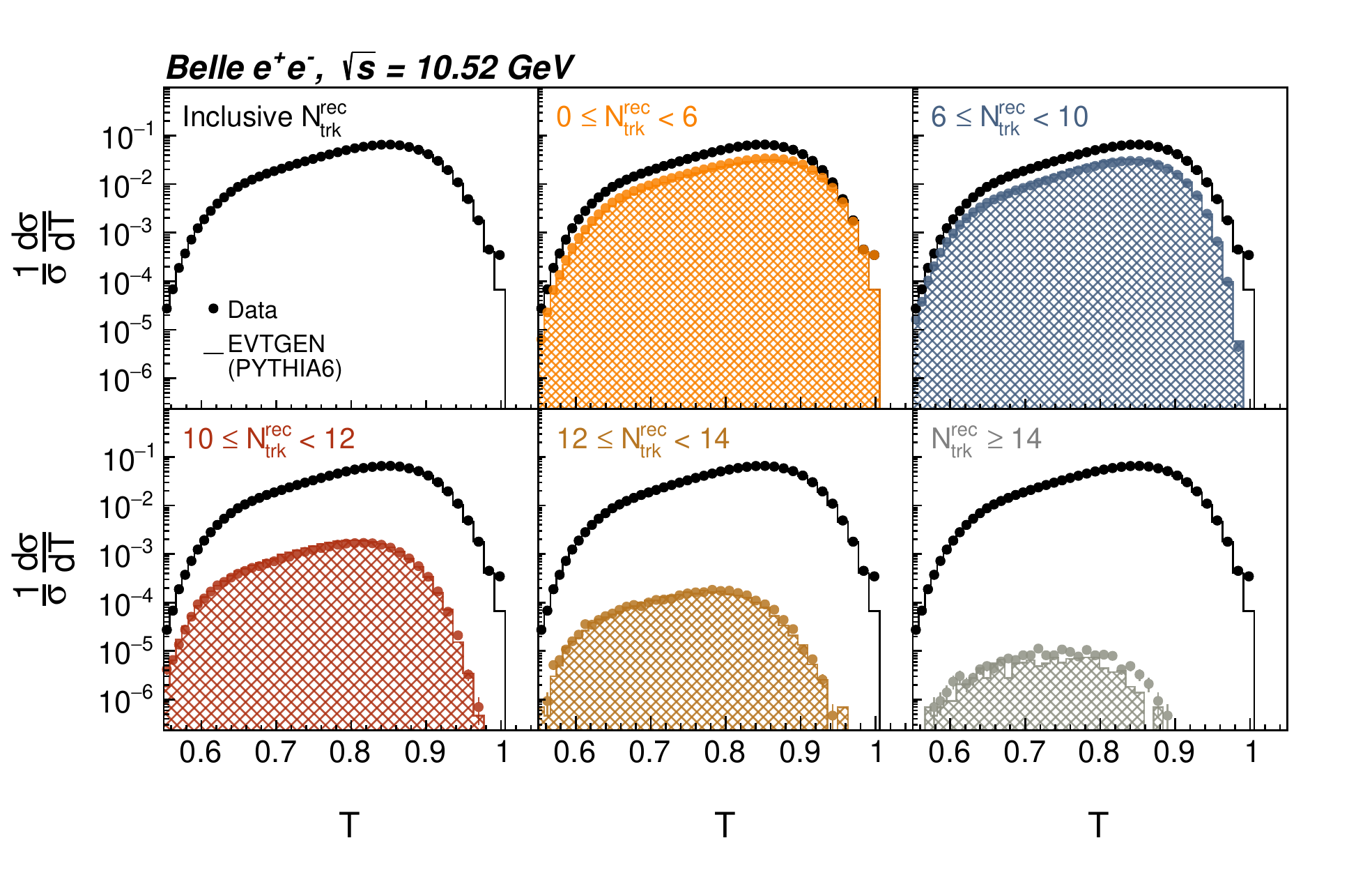}
	\caption{\conti}
    \end{subfigure}
    \begin{subfigure}[b]{\textwidth}
	\centering
	\includegraphics[width=0.8\textwidth]{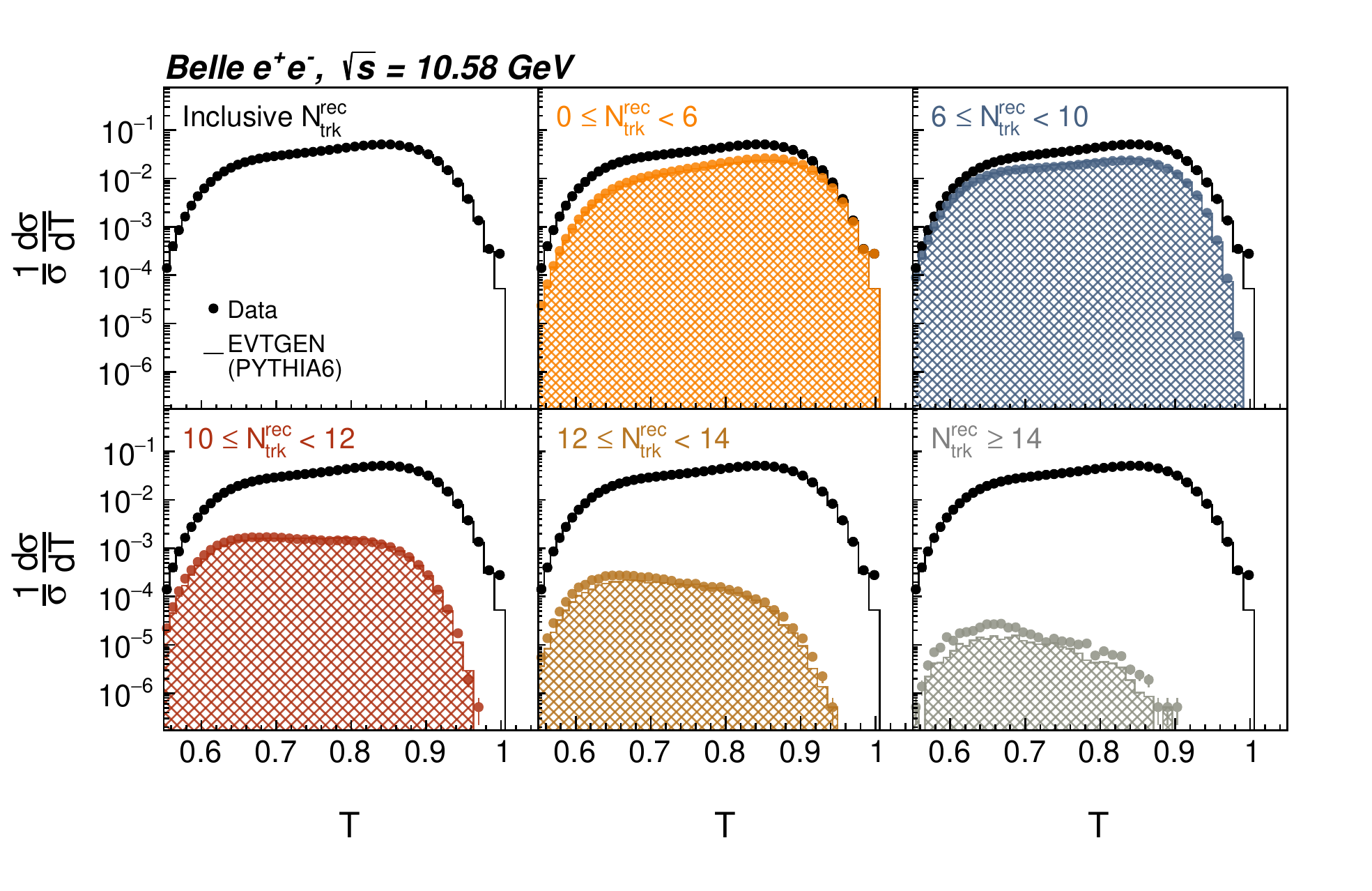}
	\caption{\onres}
    \end{subfigure}
\caption{
The thrust distributions for (a) off-resonance and (b) on-resonance samples using the same partial samples as figure~\ref{fig:NtrkOffline}. 
Results of the data (points) and the \textsc{pythia6}-based Belle MC (hatched areas) are overlaid.
The multi-panel histograms show the inclusive distributions and events with ${\rm N}_{\rm trk}^{\rm rec}$ in $[0,6)$, $[6,10)$, $[10,12)$, $[12,14)$ and $[14,\infty)$ intervals in black, gold, blue, red, brown and gray, respectively.
}
\label{fig:thrust}
\end{figure}

Since the electron and positron beam energies delivered by KEK-B are different, events are analyzed after being boosted to the \ee collision center-of-mass frame.
The event thrust axis $\hat{n}$~\cite{PhysRevLett.39.1587} is obtained through maximizing the sum of the projected particle momenta on itself in the center-of-mass frame, formulated as
\begin{equation}
T~~{\accentset{\rm max}=}~~\frac{\sum\nolimits_{i} \left| \Vec{p_i} \cdot \hat{n} \right|}{\sum\nolimits_{i} \left| \Vec{p_i} \right|},
\label{eqn:Thrust}
\end{equation}
where $T$ is the resulting thrust value and $\Vec{p_i}$ is the momentum of the $i$-th particle, with the sum extending over charged, neutral particles and the missing momentum.
Neutral particles are ECL clusters without an association from tracks' extrapolations. Requirements on the ECL acceptance and a cluster energy greater than 50--150 MeV depending on detector regions are also applied.
Detailed selections are summarized in table~\ref{tab:SelectionSummary}. 
The invisible objects, such as neutrinos, and non-reconstructed particles at the reconstruction level are taken into accounts as the missing momentum in the calculation of the event thrust, given by
\begin{equation}
\Vec{p}_{\rm MET} = -\sum\nolimits_{\rm neu, chg} \Vec{p}.
\label{eqn:MissP}
\end{equation}
Figure~\ref{fig:thrust} shows the event thrust distributions for the \conti and \onres dataset in bins of \ntrkoff, with the comparisons to the MC. The differences between data and MC are addressed by reweighting studies which are discussed in Section~\ref{sec:MCReweighting}.

\section{Two-particle correlation function in the thrust axis coordinate}
\label{sec:TwoParticleCorrelationFunction}
 
To measure two-particle correlation functions, we first perform a boost back into the $e^+e^-$ center-of-mass frame.
Two-particle correlations are then calculated by using methods previously employed in heavy ion collision and hadron collision measurements, which have been introduced in ref.~\cite{Belle:2022fvl} and formulated as
\begin{equation}
\begin{aligned}
\frac{1}{{\rm N}_{\rm trk}^{\rm corr}}\frac{d^2{\rm N}^{\rm corr}_{\rm pair}}{d\Delta\eta d\Delta\phi} &= B(0,0) \times \frac{S(\Delta\eta, \Delta\phi)}{B(\Delta\eta, \Delta\phi)},\\
\end{aligned}
\label{eqn:2PC}
\end{equation}
where ${\rm N}_{\rm trk}^{\rm corr}$ denotes the number of tracks and ${\rm N}^{\rm corr}_{\rm pair}$ the pairing yield, after efficiency correction in the event.
The correlation is analyzed in terms of the track-pair's pseudorapidity difference: $\Delta \eta = \pm (\eta_i - \eta_j)$ and azimuthal angle difference: $\Delta \phi = \pm (\phi_i - \phi_j)$. We count four entries of angular differences per track pair by assuming that the correlation function is symmetric about the origin $(\Delta \eta, \Delta \phi) = (0,0)$.
The right-hand side of the equation indicates that the correlation is obtained by factoring out the baseline background correlation function $B(\deta, \dphi)$ from the signal correlation function $S(\deta, \dphi)$, and scaled by the $B(0,0)$ factor. The signal and background correlation, formulated as
\begin{equation}
\begin{aligned}
S(\Delta\eta, \Delta\phi) &= \frac{1}{{\rm N}_{\rm trk}^{\rm corr}}\frac{d^2{\rm N}^{\rm same}} {d\Delta\eta d\Delta\phi},\\
B(\Delta\eta, \Delta\phi) &= \frac{1}{{\rm N}_{\rm trk}^{\rm corr}}\frac{d^2{\rm N}^{\rm mix}}{d\Delta\eta d\Delta\phi},
\end{aligned}
\label{eqn:2PCSigBkgFunctions}
\end{equation}
count the per-trigger-particle~\footnote{To describe the pairing between two particles $i$-th and $j$-th, this paper uses the term ``trigger particle'' for the $i$-th particle and the ``associated particle'' for the $j$-th one for convenience. In other two-particle correlation analyses, there can be different selection criteria for trigger and associated particles, but in this paper, both are from the same collection if not further specified.} pairing yields within the same event and with the ``mixed event''~\cite{CMS:2011cqy,Chatrchyan:2012wg,CMS:2012qk,CMS:2013jlh,CMS:2014und}. 
A mixed event in this analysis is formed by pairing events with similar ${\rm N}_{\rm trk}^{\rm rec}$. In the analysis, trigger particles from a signal event are matched with associated tracks from three other events separately, in order to increase the statistical accuracy of the mixed event sample.
The pairing yield in the background correlation function (${\rm N}^{\rm mix}$) is obtained by counting the number of associated tracks in the mixed event for each trigger particle in the signal event. Since the associated track and the trigger track come from different events, the pair is totally uncorrelated. 
The scale factor $B(0,0)$ is the bin value at the origin of the background correlation function, accounting for the normalization of the artificially constructed background function.

In this work, we label the reference coordinate system of this classic two-particle correlation formulation, referring to the beam pipe as its $z$-axis for the calculation of tracks' $\eta$ and $\phi$ coordinates, as the ``beam axis coordinates.''
However, the dominant energy flow of the \ee collision events follow the directions of the out-going quark and anti-quark produced in the electron-position annihilation, which differ event-by-event.
Consequently, in the fixed beam axis coordinate system, such dominant dijet-like topology can be smeared out,  affecting the observation of any smaller correlation structure.
More specifically, the leading and the subleading jets, which make up the main back-to-back correlation feature in  $\ee$ collision data, will contribute with arbitrary pseudorapidity differences (i.e., not necessarily at large $\Delta \eta$) depending on the outgoing dijet direction.

In addition to the conventional beam axis analysis construction, this paper documents the methodology of novel two-particle correlations measured in the ``thrust axis coordinates.''
This analysis method is devised initially by ref.~\cite{Badea:2019vey} for \ee collision data, taking the event thrust axis, which approximates the outgoing direction of $\ep\en$ collision events at the parton level, as the reference $z$-axis.
The difference in the thrust axis analysis is that tracks' coordinates ($\eta, \phi$) are recalculated with respect to the new reference $z$-axis, the thrust axis $\hat{n}$.
The reference axis for $\phi=0$ (or new reference $x$-axis) is assigned with $\hat{n}\times(\hat{n}\times\hat{z})$, where $\hat{n}$ and $\hat{z}$ are the unit vectors of the event thrust and the beam direction. This defines a new set of coordinate values in the thrust axis reference frame ($\eta_T, \phi_T$), which can be used to calculate the angle separations and to form the signal and background correlation distributions in eq.~(\ref{eqn:2PC}).

For the background correlation function calculation, we use the same event-mixing method as the typical beam axis analysis by combining  different events to form a mixed event, which is then used to account for the random pairing baseline correlation in the thrust axis analysis.
However, since the thrust axis of each paired event is randomly orientated, 
a thrust-mixing reweighting correction on particles' thrust-referenced angular distribution ($\eta_T, \phi_T$) is adopted. Such a correction is needed since the variation of the thrust axis direction causes detector acceptance in the thrust coordinates to vary on an event-by-event basis.
The reweighting factor is obtained from a histogram division of the physical spectrum by the mixed-event spectrum, written out as
\begin{equation}
m(|\eta_T|, \phi_T, {\rm N}_{\rm trk}^{\rm rec}) = \left[\frac{d^2{\rm N}^{\rm phys}}{d|\eta_T| d\phi_T}/\frac{d^2{\rm N}^{\rm mix}}{d|\eta_T| d\phi_T}\right]_{{\rm N}_{\rm trk}^{\rm rec}},
\label{eqn:ThrustMixingCorrection}
\end{equation}
which can force the mixed-event single-particle angular distribution to mimic the physical one. 
The reweighting factor is further applied as track weights on the background two-particle correlation function $B(\deta,\dphi)$.
This ensures that the event topology of mixed events conform to that of physical events, and leaves the pairing of the trigger particle with tracks from the mixed event uncorrelated at the same time.

Since the particle's pseudorapidity evolves non-linearly towards divergence when getting close to the reference axis,
the two-particle correlation function observable defined on the \deta-\dphi plane monitors mainly final states traveling in the transverse direction to the reference axis.
In other words, particle pairs with very large $\eta$ values  are beyond the scope of interest for the \deta window.
Thus, when measuring with the correlation defined with respect to the beam axis for heavy ion collisions, one is looking at the mid-rapidity region where final states expand from a possible QGP, while backgrounds from beam remnants in the extreme forward and backward regions are out of the pseudorapidity coverage of the observable defined.
For the thrust axis coordinates, the mid-rapidity region probes the physics of soft emission while the leading-order quark-initiated jet correlations with large pseudorapidity differences are omitted.

A projection of correlations to the particle pairs' azimuthal angle difference \dphi is also studied, given by integrating and averaging two-dimensional correlation functions over $\Delta\eta$ from $\Delta\eta_{\rm min}$ to $\Delta\eta_{\rm max}$,
\begin{equation}
Y(\Delta\phi) = \frac{1}{{\rm N}_{\rm trk}^{\rm corr}}\frac{d{\rm N}^{\rm corr}_{\rm pair}}{d\Delta\phi}= \frac{1}{\Delta\eta_{\rm max}-\Delta\eta_{\rm min}}\int\limits_{\Delta\eta_{\rm min}}^{\Delta\eta_{\rm max}} \frac{1}{{\rm N}_{\rm trk}^{\rm corr}}\frac{d^2{\rm N}^{\rm corr}_{\rm pair}}{d\Delta\eta d\Delta\phi}d\Delta\eta.
\label{eqn:DeltaPhiAssociatedYield}
\end{equation}
We explore this azimuthal differential associated yield for the full range of \dphi and in three $\Delta\eta$ regions: short range ($0 \le |\Delta\eta| < 1$), middle range ($1 \le |\Delta\eta| < 1.5$) and long range ($1.5 \le |\Delta\eta| < 3.0$). The long-range azimuthal differential associated yield is denoted as $Y_l(\Delta \phi)$ in the following content.

\section{Corrections}
\label{sec:Corrections}

\subsection{Tracking efficiency correction}
\label{sec:EfficiencyCorrection}

With the Belle MC sample as introduced in section~\ref{sec:Sample}, we study the reconstruction effects of nonuniform detection efficiency and misreconstruction bias.
A reweighting factor is applied to the reconstructed tracks by the inverse of the tracking efficiency, bringing the reconstructed-track spectra closer to reflecting the geometrical acceptance of the detector. The efficiency is given by
\begin{equation}
\varepsilon(p_{\rm T}, \theta, \phi, {\rm N}_{\rm trk}^{\rm rec}) = \left[\frac{d^3 {\rm N}^{\rm reco}}{dp_{\rm T} d\theta d\phi}/\frac{d^3 {\rm N}^{\rm gen}}{{dp_{\rm T} d\theta d\phi}}\right]_{{\rm N}_{\rm trk}^{\rm rec}},
\label{eqn:Efficiency}
\end{equation}
where ${\rm N}^{\rm gen}$ denotes the number of primary charged particles (as defined in table~\ref{tab:SelectionSummary}) counted at the generator level, and ${\rm N}^{\rm reco}$ denotes those at the reconstruction level. The tracking efficiency correction addresses reconstruction effects such as the inclusion of secondary particles and detector effects. 

\subsection{MC reweighting}
\label{sec:MCReweighting}

There are potential differences between data and simulation: for instance, some processes are not taken into account, or the proportion of each process is not estimated accurately, etc.
The MC sample used to model the reconstruction effects is thus reweighted to correct for the imperfection in MC simulation.
A ``MC reweighting factor $r$'' is applied as the event weight to adjust the discrepancy. This factor is obtained by a histogram ratio, 
\begin{equation}
r({\rm N}_{\rm trk}, T) =  \langle \frac{d^2 {\rm N}^{\rm data}}{d{\rm N}_{\rm trk} dT}  \rangle / 
\langle \frac{d^2 {\rm N}^{\rm MC}}{d{\rm N}_{\rm trk}dT} \rangle,
\label{eqn:ReweightingFactor}
\end{equation}
where the numerator and the denominator are the data and MC event multiplicity (${\rm N}_{\rm trk}$) and thrust ($T$) distributions normalized by number of events, respectively.
For events with the number of tracks less than $12$, the ratios are close to 1 ($0.98$--$1.08$), while the ratios are larger in high-multiplicity
events, shooting up to $1.14$--$1.34$.

The MC reweighting factor $r$ is incorporated into the calculation of the efficiency correction factor (see eq.~(\ref{eqn:Efficiency})) to derive the new factor $\varepsilon'$, given by
\begin{equation}
\varepsilon'(p_{\rm T}, \theta, \phi, {\rm N}_{\rm trk}^{\rm rec}) = \left[\frac{d^3[r({\rm N}_{\rm trk}, T) {\rm N}^{\rm reco}]}{dp_{\rm T} d\theta d\phi}/\frac{d^3[r({\rm N}_{\rm trk}, T) {\rm N}^{\rm gen}]}{{dp_{\rm T} d\theta d\phi}}\right]_{{\rm N}_{\rm trk}^{\rm rec}}.
\label{eqn:EfficiencyAfterReweighting}
\end{equation}

\subsection{Bin-size effect correction --- \bzz normalization extrapolation and long-range correlations scaling}
\label{sec:B00Ext}

In the calculation of two-particle correlations (see eq.~(\ref{eqn:2PC})), we account for the normalization factor \bzz by extrapolation.
This procedure is used to resolve any artificial bin-width effect\footnote{
The correlation function taken as the number of pairing yields per $\deta(\dphi)$ is an expression of the density function ($d^2N/\text{(bin width)}^2$), hence there is no trivial bin-width dependency.
However, different $\deta(\dphi)$ window sizes also affect the pairing yields average; this second-order bin size effect is estimated in this work.
} in the conventional method, where \bzz is given by the yield in the zeroth bin.
The effects introduced by using finite-bin histogramming to approximate the correlation function are studied independently for $\Delta\eta$ and $\Delta\phi$.
The impact of the variation of the $\Delta\phi$ bin width is checked to have a small effect on the $B(0,0)$ value compared to that caused by the variation of the $\Delta\eta$ bin width.
Therefore, we focus only on the correction of the $\Delta\eta$ bin choice dependency on the raw $B(0,0)$ value, leaving the uncertainty regarding the $\Delta\phi$ bin choice as one of the sources of systematic uncertainty. 

The trend of the \bzz value as a function of the \deta bin-width configuration is modeled with a second-order polynomial asymptotically approaching a constant as the \deta bin width goes to zero. We use as the extrapolation of \bzz, the value at the zero \deta bin width.

For the beam axis analysis, a residual bin-size effect on the normalization of the long-range azimuthal differential yield $Y_l(\dphi)$ is found.
Hence, we calibrate the long-range correlation magnitude by the same approach as that to correct the \bzz value.
The final beam-axis $Y_l(\dphi)$ is multiplied by a scaling factor  $\frac{\int Y_{l}(\Delta\phi)d\Delta\phi]_{ext}}{\int Y_{l}(\Delta\phi)d\Delta\phi]_{hist}}$,
where the numerator is the total associated yield in the long-range region using a second-order-polynomial extrapolation, and the denominator is the raw histogram value.

\subsection{Residual MC correction}

After the above corrections, remaining possible reconstruction effects are handled with the bin-by-bin correction method~\cite{Choudalakis:2011rr}.
The correction factor is derived from the ratio of MC correlation functions at the reconstruction and generator level as ${\rm C}(\Delta \phi) = \frac{Y(\Delta \phi)_{{\rm gen}, i_g}}{Y(\Delta \phi)_{{\rm reco}, i_r}}$, where
indices $i_g$ and $i_r$ are ${\rm N}_{\rm trk}^{\rm rec}$ bins counted at the generator and reconstruction level, respectively. The factors correct $2$-$5\%$ discrepancies between the reconstruction and generator level.
Final data correlation results are obtained from the multiplication of the original correlation function with the bin-by-bin correction factor.

\section{Results}
\label{sec:Rst}
 
\begin{figure}[ht]
\centering
\includegraphics[width=\textwidth]{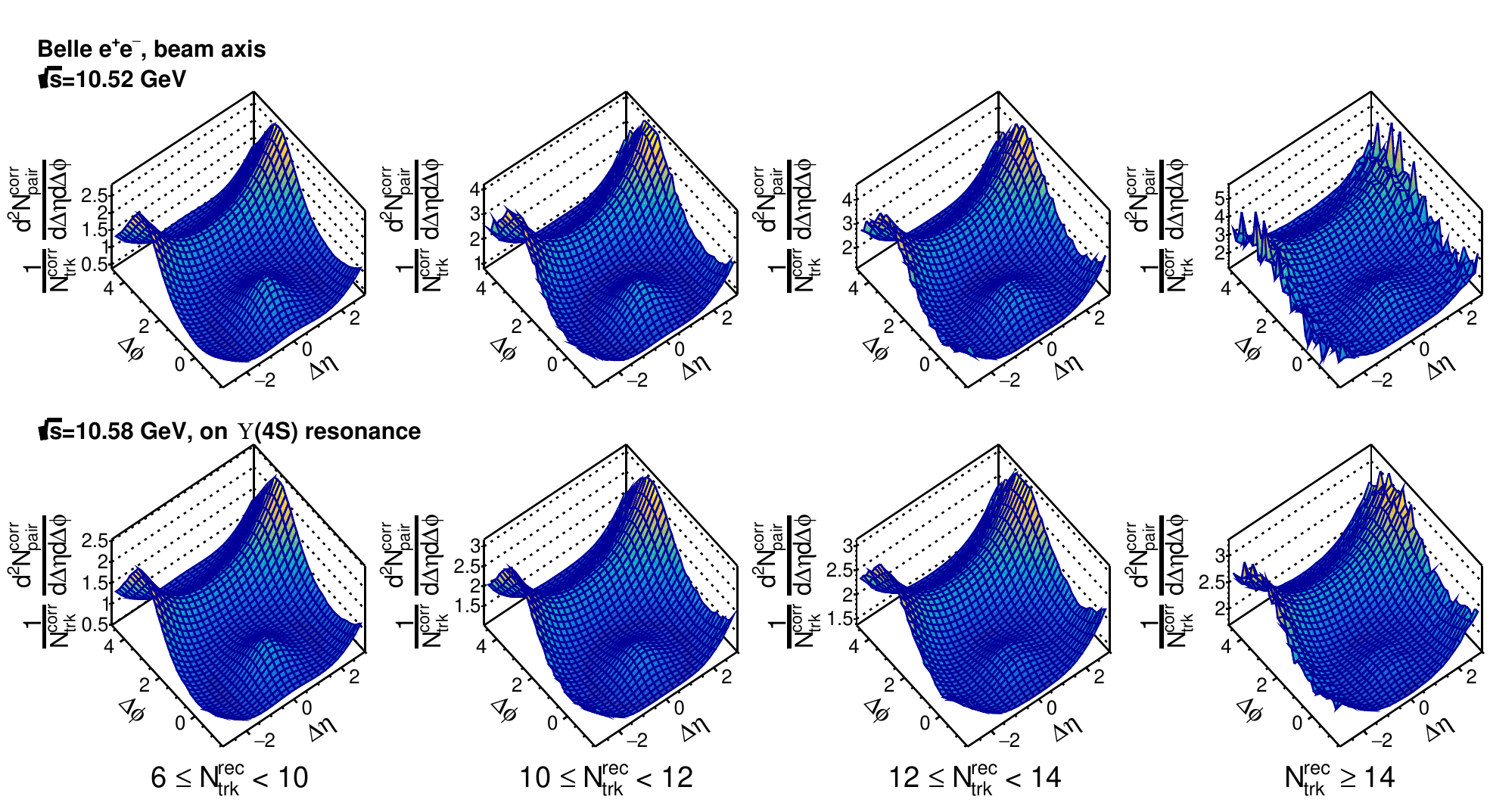}
\caption{Beam-axis two-particle correlation functions for the \conti (top) and \onres (bottom) data versus the reconstructed-track multiplicity ${\rm N}_{\rm trk}^{\rm rec}$ in $[0,6)$, $[6,10)$, $[10,12)$ and $[14,\infty)$ intervals from left to right.}
\label{fig:Data2PC_beam}
\end{figure}

Partial results of the two-particle correlations measured using the \conti dataset are published in ref.~\cite{Belle:2022fvl}.
In figure~\ref{fig:Data2PC_beam}, beam-axis two-particle correlations for the \conti and \onres data are shown as a function of \ntrkoff. 
The correlation structure in view of the beam axis analysis can be understood in the following decomposition.
The {\it near-side peak} correlation, residing near $(\Delta\eta, \Delta\phi)=(0,0)$, has contributions mainly from track pairs that originate in the same jet. 
The {\it away-side} correlation, elongated along $\Delta\phi\approx\pi$, results from  back-to-back momentum balancing. 
This correlation in $e^+e^-$ annihilation has a feature that tilts upwards to large \deta. 
Since the annihilation products of the \ee collision do not happen to align with the direction of the beamline, this increases the probability of finding back-to-back emitted final states at larger pseudorapidity difference.
The typical {\it ridge} correlation is the structure with $\dphi\approx0$ and persisting over a wide range of \deta~\cite{Abelev:2012ola}. This is not visually observed in the Belle data.

\begin{figure}[ht]
\centering
\includegraphics[width=\textwidth]{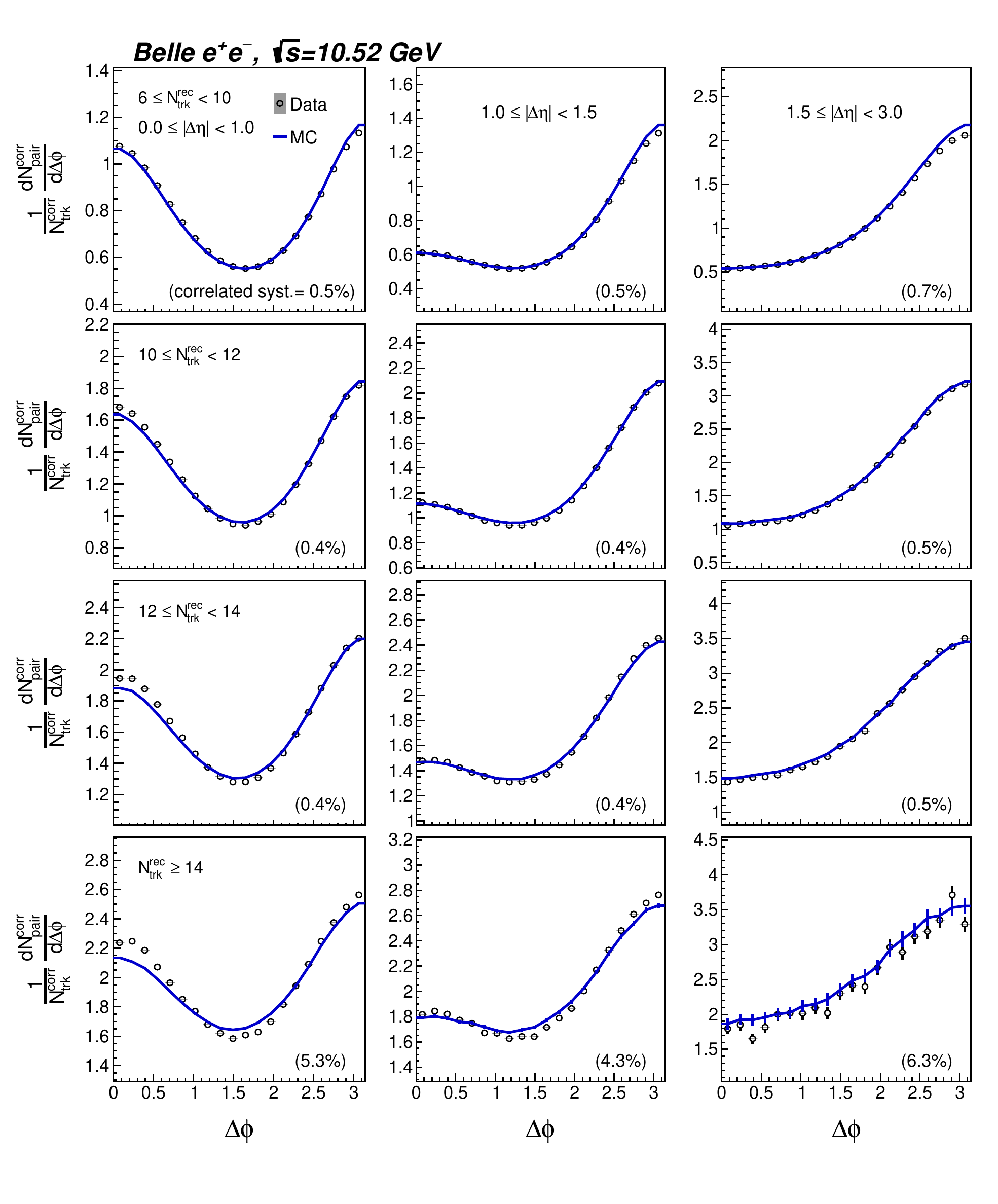}
\caption{Beam-axis azimuthal differential associated yields as a function of the reconstructed-track multiplicity \ntrkoff for the off-resonance sample in the short-range ($0 \le |\Delta\eta| < 1$), middle-range ($1 \le |\Delta\eta| < 1.5$) and long-range ($1.5 \le |\Delta\eta| < 3.0$) regions in columns from left to right, respectively.}
\label{fig:ContiBeam2PCProj}
\end{figure}

\begin{figure}[ht]
\centering
\includegraphics[width=\textwidth]{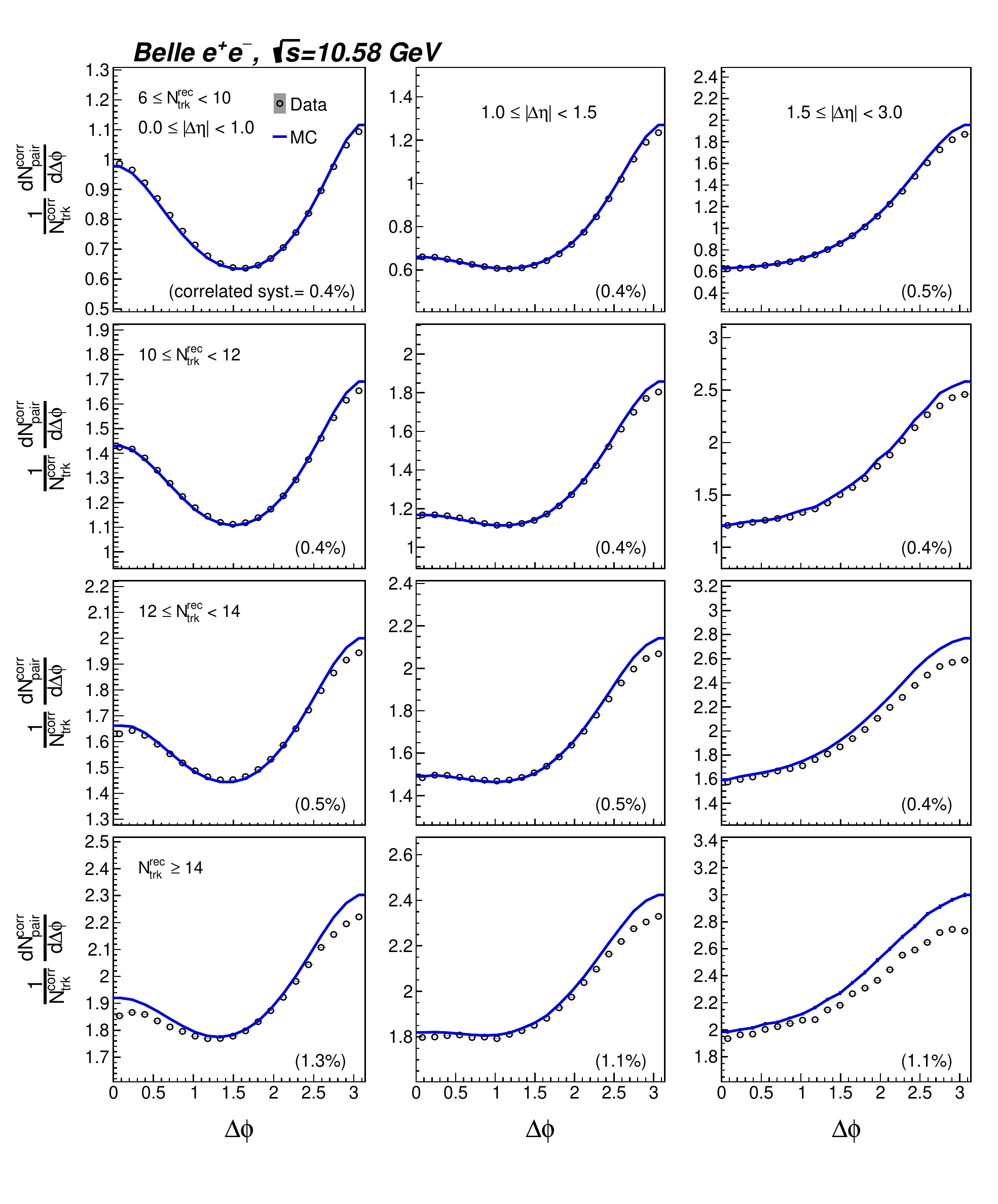}
\caption{Beam-axis azimuthal differential associated yields as a function of the reconstructed-track multiplicity \ntrkoff for the on-resonance sample in the short-range ($0 \le |\Delta\eta| < 1$), middle-range ($1 \le |\Delta\eta| < 1.5$) and long-range ($1.5 \le |\Delta\eta| < 3.0$) regions in columns from left to right, respectively.}
\label{fig:OnresBeam2PCProj}
\end{figure}

Zoom-in displays of beam-axis correlation functions in different \deta ranges are shown by azimuthal differential associated yield distributions in figures~\ref{fig:ContiBeam2PCProj} and~\ref{fig:OnresBeam2PCProj} with the \conti and \onres data, respectively. Results for data (black points with error bars) and MC (blue curves with error bars) are overlaid in the short range (left), middle range (middle) and long range (right) region for comparison. Plot labels indicate the correlated systematic uncertainties over \dphi bins; uncorrelated systematic uncertainties depending on the \dphi value are plotted with gray boxes.
There are some discrepancies in the near-side ($\dphi\approx0$) and the away-side ($\dphi\approx\pi$) peak values between the \onres data and MC results, while for the $\ee \to \qq$ \conti sample, the MC provides a better match with the data.

\begin{figure}[ht]
\centering
\includegraphics[width=\textwidth]{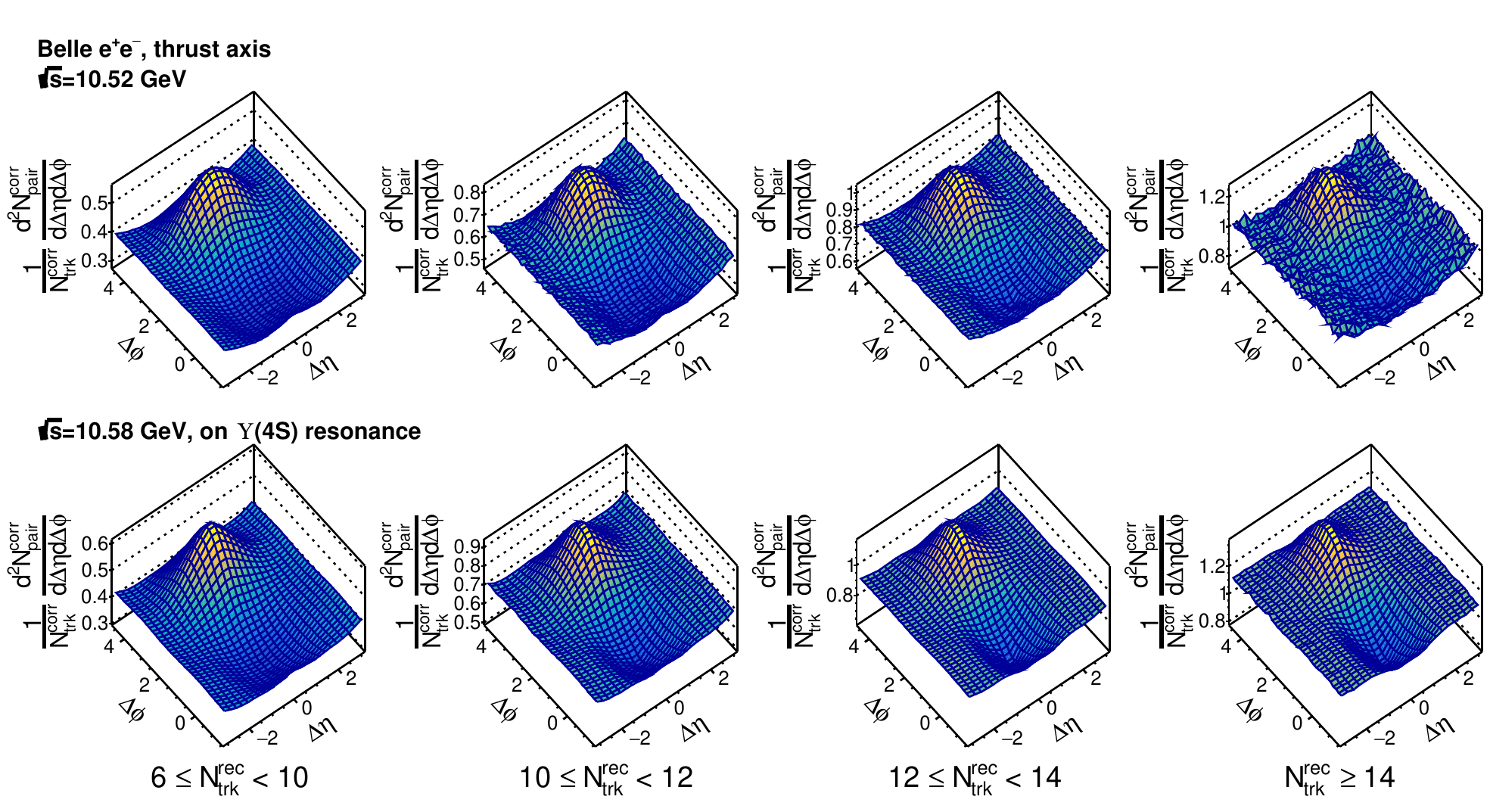}
\caption{Thrust-axis two-particle correlation functions for the \conti (top) and \onres (bottom) data versus the reconstructed-track multiplicity ${\rm N}_{\rm trk}^{\rm rec}$ in $[0,6)$, $[6,10)$, $[10,12)$ and $[14,\infty)$ intervals from left to right.}
\label{fig:Data2PC}
\end{figure}

In figure~\ref{fig:Data2PC}, thrust-axis two-particle correlations for the \conti and \onres data are shown as a function of the reconstructed-track multiplicity.
The magnitude of correlations decreases significantly compared with that of the beam axis analysis, due to the exclusion of the dominant back-to-back correlation at the extreme pseudorapidity difference.
The main correlation structure is the bump at the away-side region; compared with the \ee collisions at the $Z$-pole energy~\cite{Badea:2019vey}, a sizable near-side peak correlation is lacking. 
We investigate this latter effect in section~\ref{sec:GeneratorStudy} by simulations, finding that the near-side peak correlation of \ee annihilation depends strongly on the collision energy.
In general, qualitatively similar results are observed for the \conti and \onres data.

\begin{figure}[ht]
\centering
\includegraphics[width=\textwidth]{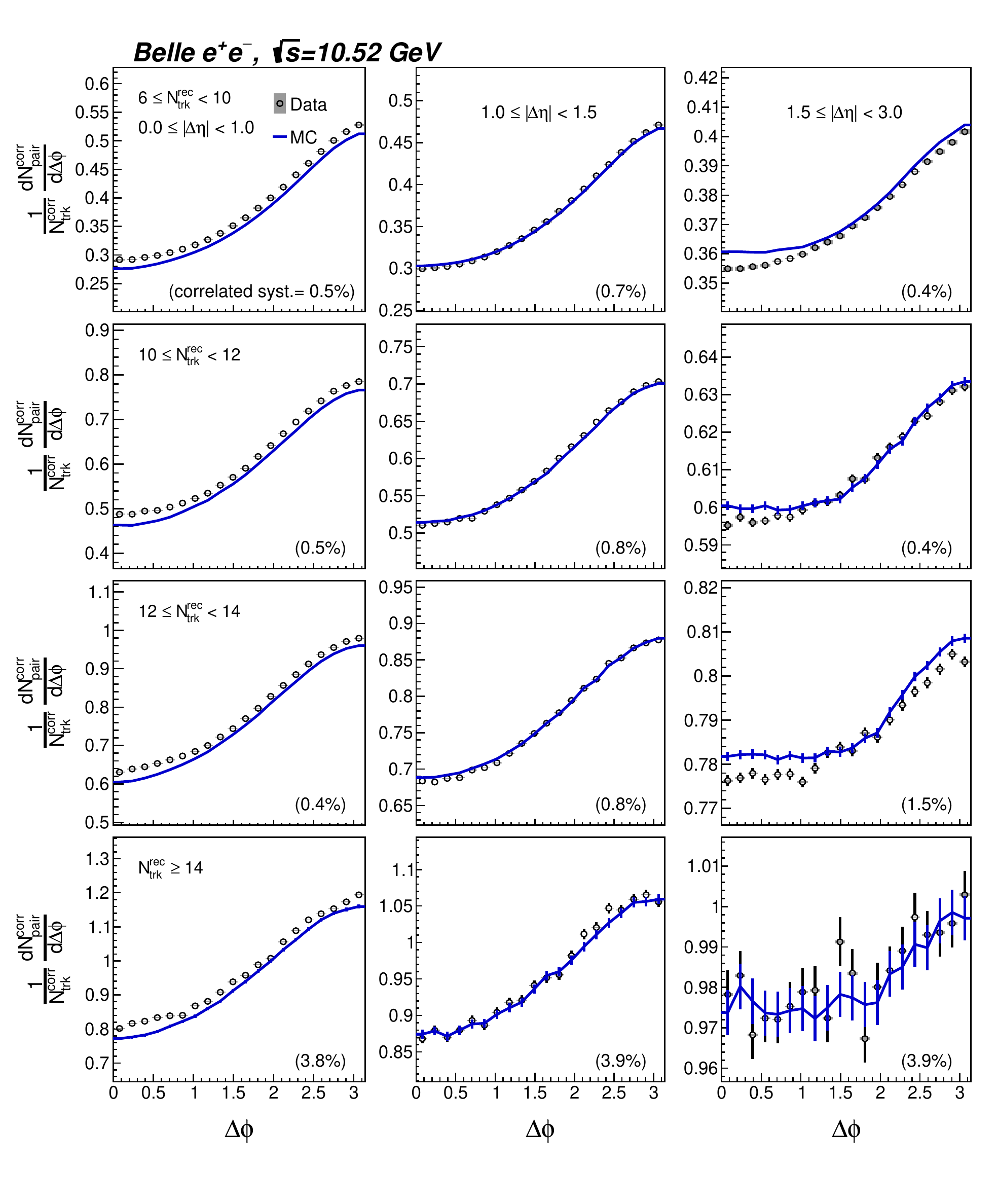}
\caption{Thrust-axis azimuthal differential associated yields as a function of the reconstructed-track multiplicity \ntrkoff for the off-resonance sample in the short-range ($0 \le |\Delta\eta| < 1$), middle-range ($1 \le |\Delta\eta| < 1.5$) and long-range ($1.5 \le |\Delta\eta| < 3.0$) regions in columns from left to right, respectively.}
\label{fig:ContiThrust2PCProj}
\end{figure}

\begin{figure}[ht]
\centering
\includegraphics[width=\textwidth]{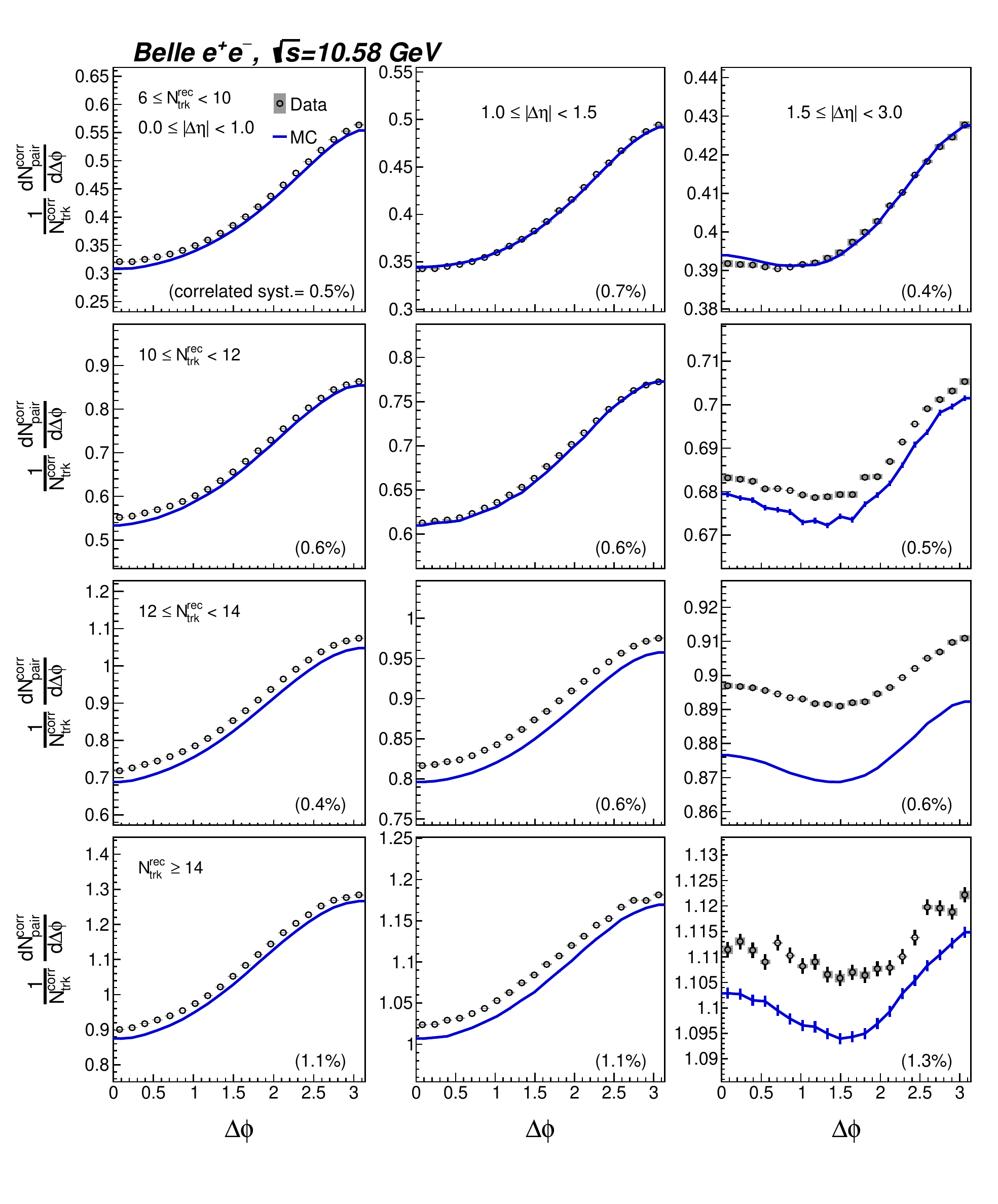}
\caption{Thrust-axis azimuthal differential associated yields as a function of the reconstructed-track multiplicity \ntrkoff for the on-resonance sample in the short-range ($0 \le |\Delta\eta| < 1$), middle-range ($1 \le |\Delta\eta| < 1.5$) and long-range ($1.5 \le |\Delta\eta| < 3.0$) regions in columns from left to right, respectively.}
\label{fig:OnresThrust2PCProj}
\end{figure}

Figures~\ref{fig:ContiThrust2PCProj} and~\ref{fig:OnresThrust2PCProj} show details of the data and MC comparison for thrust-axis coordinate azimuthal differential associated yields with the \conti and \onres data, respectively.
In the small \deta region, there are mild-level magnitude differences between data and MC correlation. In the long range region, a qualitatively better agreement in correlation function shapes between the data and MC is observed for \conti results; however,  there is up to a 5\% larger discrepancy seen in the  correlation function magnitude for the \onres sample.
For the search of the ridge signal, the $\ee \to \qq$ \conti dataset reveals no evidence of special enhancement, as reported by the measurement with ALEPH data~\cite{Badea:2019vey} and the recent Belle Letter~\cite{Belle:2022fvl}.
For the \onres data, it is worth noting that there is an enhanced long-range near-side correlation. The enhancement is visible for events with multiplicity greater than 10 in the rightmost columns of figure~\ref{fig:OnresThrust2PCProj}.
However, it doesn't resemble the typical ridge structure observed widely in the large collision systems, which is an enhanced correlation region over a broad range of \deta with small \dphi. We also observe a similar enhancement with MC, and provide further investigation of the enhancement using MC in section~\ref{sec:GeneratorStudy}.

\clearpage

The systematic uncertainties due to selection and correction operations are calculated with respect to the long-range associated yield $Y_l(\Delta \phi)$, and listed as percentages in tables~\ref{tab:Systematics_beam} and~\ref{tab:Systematics} for the beam-axis and thrust-axis analyses, respectively. 
In the tables, correlated systematic uncertainties for all \dphi bins are reported. The associated yield variation from each systematic source discussed below is observed in general to be uniform along \dphi, while a mild level of non-uniform systematic uncertainty on the associated yield is observed in the variation of the thrust mixing reweighting factor parametrization (see eq.~(\ref{eqn:ThrustMixingCorrection})). The effect is estimated to be of order $0.01\%$. 
The uncorrelated uncertainties depending on \dphi values are shown in the azimuthal differential associated yield plots (figures~\ref{fig:ContiBeam2PCProj},~\ref{fig:OnresBeam2PCProj},~\ref{fig:ContiThrust2PCProj} and~\ref{fig:OnresThrust2PCProj}). 

\begin{table}[ht] 
\footnotesize
\caption{Beam axis analysis systematic uncertainties for the \conti and \onres data as a function of the reconstructed-track multiplicity \ntrkoff. All values are reported as percentages of the long-range differential associated yield.}
\begin{center}
\begin{tabularx}{\textwidth}{C|CCCCCCC}
\hline\hline
\hspace{0.2cm}${\rm N}_{\rm trk}^{\rm rec}$ \hspace{0.2cm} & Primary particle selection & Tracking efficiency  & Event selection & $\Delta\phi$ bin width     & $B(0,0)$ extrapolation  &  Long-range scaling  &  MC reweighting \\
\hline
\multicolumn{8}{c}{\conti}\\
\hline 
$[6,10)$ & 0.42 & 0.35 & 0.38 & 0.01 & 0.01 & 0.01 & 0.01\\
\hline
$[10,12)$ & 0.30 & 0.35 & $<$0.01 & $<$0.01 & 0.03 & 0.04 & 0.03\\
\hline
$[12,14)$ & 0.39 & 0.35 & $<$0.01 & 0.02 & 0.04 & 0.08 & 0.04\\
\hline
$[14,\infty)$ & 6.27 & 0.35 & $<$0.01 & 0.02 & 0.14 & 0.30 & 0.20\\
\hline
\multicolumn{8}{c}{\onres}\\
\hline 
$[6,10)$ & 0.25 & 0.35 & 0.25 & 0.02 & 0.01 & 0.01 & 0.02\\
\hline
$[10,12)$ & 0.02 & 0.35 & $<$0.01 & $<$0.01 & 0.02 & 0.03 & $<$0.01\\
\hline
$[12,14)$ & 0.04 & 0.35 & $<$0.01 & 0.03 & 0.01 & 0.01 & 0.04\\
\hline
$[14,\infty)$ & 1.02 & 0.35 & $<$0.01 & 0.02 & 0.02 & 0.05 & 0.08\\
\hline
\hline
\end{tabularx} 
\label{tab:Systematics_beam}
\end{center}
\end{table}

\begin{table}[ht] 
\footnotesize
\setlength\extrarowheight{1.5pt} 
\caption{Thrust axis analysis systematic uncertainties for the \conti and \onres data as a function of the reconstructed-track multiplicity \ntrkoff. All values are reported as percentages of the long-range differential associated yield.}
\begin{center}
\begin{tabularx}{\textwidth}{C|CCCCCCC}
\hline\hline
\hspace{0.2cm}${\rm N}_{\rm trk}^{\rm rec}$ \hspace{0.2cm} & Primary particle selection   &  Tracking efficiency     & Event selection  & $\Delta\phi$ bin width     & $B(0,0)$ extrapolation  &  MC reweighting & Thrust mixing reweighting  \\
\hline
\multicolumn{8}{c}{\conti}\\
\hline
$[6,10)$ & 0.15 & 0.35 & 0.04 & 0.05 & 0.02 & $<$0.01 &
 0.16\\
\hline
$[10,12)$ & 0.01 & 0.35 & 0.26 & 0.06 & 0.04 & 0.02 
& 0.03\\
\hline
$[12,14)$ & 0.71 & 0.35 & 1.32 & 0.03 & 0.06 & 0.02 
& 0.09\\
\hline
$[14,\infty)$ & 3.81 & 0.35 & 0.15 & 0.23 & 0.21 & 0.12
 & 0.25\\
\hline
\multicolumn{8}{c}{\onres}\\
\hline
$[6,10)$ & 0.03 & 0.35 & 0.07 & 0.04 & 0.01 & 0.03 &
 $<$0.01\\
\hline
$[10,12)$ & 0.27 & 0.35 & 0.06 & 0.05 & 0.03 & 0.01 
& 0.04\\
\hline
$[12,14)$ & 0.47 & 0.35 & 0.02 & 0.02 & 0.01 & 0.03 
& 0.02\\
\hline
$[14,\infty)$ & 0.12 & 0.35 & 1.13 & 0.01 & 0.03 & 0.06
 & 0.45\\
\hline\hline
\end{tabularx} 
\label{tab:Systematics}
\end{center}
\end{table}

The primary particle selection systematic uncertainty is estimated by changing the generator definition of the proper lifetime requirement $\tau<1$ cm/$c$ to the vertex requirement $V_{r}<1$ cm.
This variation of different truth-level definitions enters in the correction factor calculation for the tracking efficiency.
A 0.35\% uncertainty on the reconstruction efficiency for high-momentum tracks with $p_{\rm T}>200$ \mevc~\cite{BaBar:2014omp}  propagates directly to the calculation of the per-trigger-particle associated yield magnitude.
The sample in this analysis passes hadronic event selections~\cite{Belle:2001jqo}, which requires the energy sum ($E_{\rm sum}$) in the ECL to be greater than $0.18\sqrt{s}$. 
We examine the systematic effect of hadronic event selection by tightening it to $0.23\sqrt{s}$. The variation of the event selection has an impact in the low-multiplicity region primarily, while the effect is not significant for high multiplicity events.

We correct for the histogramming bin-size effect of the two-particle correlation function by extrapolating to the zero \deta bin width discussed in section~\ref{sec:B00Ext} and consider the \dphi bin choice as a source of systematic uncertainties. 
The nominal \dphi bin width is 0.157; we recalculate with an alternative bin choice, corresponding to a bin width of 0.174. Either are enough to describe the correlation function with perfect granularity.
The resulting difference is quoted as one of the systematics.
We only use this alternative configuration to evaluate the \dphi bin-width systematics, and assume similar changes in the case for decreasing bin-width, given that the size of this uncertainty is roughly of order $0.01\%$ and therefore small in comparison to the dominating uncertainties.
For the $B(0,0)$ normalization factor, we calibrate the bin-size effect by extrapolation with a second-order fitting curve. The fitting error is propagated to the $Y_l(\dphi)$ and is considered as one of the systematic uncertainties. The uncertainty on the scaling factor of the long-range correlation magnitude is also given by the propagation of the fitting error.

The systematic uncertainties of reweighting factors are studied by changing their parametrization. 
For the MC reweighting factor (see eq.~(\ref{eqn:ReweightingFactor})), we change the dependence from the event multiplicity and thrust to multiplicity only;
for the thrust mixing reweighting factor (see eq.~(\ref{eqn:ThrustMixingCorrection})), we change the dependence from ($|\eta|$, $\phi$) to $|\eta|$ only.
The corresponding difference on the correlation observable is quoted as the systematic uncertainty.
The systematic uncertainties on the residual MC correction factors are assessed by changing the correction factors' parametrization to two dimensions $(\Delta \eta, \Delta \phi)$; hence, the alternative correction factors are determined using the ratio of two-dimensional correlation functions. 
Uncertainties in the correction factors are found to be no greater than $\mathcal{O}(0.01\%)$, which is a negligible effect compared with other sources of systematics. 

Systematic uncertainties from selections (including primary particles, tracks and the hadronic events), totalling from approximately $0.4 \%$--$6.3 \%$, contribute as dominant systematic uncertainties in this measurement.
Larger uncertainties appear in high-multiplicity bins, which are due to the limitation in statistics to acquire the precise track efficiency correction factor, the MC reweighting factor and the thrust-mixing reweighting factor. 
In the beam axis analysis, the variation of the event selection criteria has little effect on the change of associated yields in events with multiplicity greater than 10, while in the thrust axis analysis, since an additional operation of the thrust-mixing reweighting is applied, the uncertainties are greater.
Other uncertainties are comparably small, contributing from $0.01\%$ up to $0.5\%$ in the difference of long-range associated yields.

A common measure for the structure of the associated yield function uses the ``zero yield at minimum'' (ZYAM) method~\cite{Ajitanand:2005jj}.
The minimum of the \ydphi associated yield is shifted to the zero yield after a flat correlation contribution, denoted as a constant $C_{\rm ZYAM}$, is subtracted off.
Since correlation functions are symmetric about $\Delta\phi=0$, we consider three even-function fit templates
\begin{equation}
\def\arraystretch{1.}
f(\Delta \phi) = \left\{
\begin{array}{l}
v_0 + 2 \sum\limits_{n=1}^3 v_n \cos(n\Delta \phi),\\
a_0 + a_1 (\Delta \phi)^2 + a_2 \cos(2\Delta \phi),\\
a_0 + a_1 (\Delta \phi)^2 + a_2 (\Delta \phi)^4,
\end{array} \right.
\label{eqn:ZYAMFuncs}
\end{equation}
for the determination of the constant $C_{\rm ZYAM}$ correlation and the $x$ coordinate of the minimum ($\Delta\phi_{\rm min}$).
Thereafter, the ridge yields are quantified by integrating ZYAM-subtracted azimuthal differential associated yields over the long-range near-side ($0 \le \dphi \le \dphi_{\rm min}$) region,
\begin{equation}
Y_{\rm ridge} = \int\limits_{0}^{\Delta\phi_{\rm min}} [ Y_l(\Delta\phi) - C_{\rm ZYAM} ] d\Delta\phi.
\label{eqn:RidgeYield}
\end{equation}

\begin{figure}[ht]
\centering
\includegraphics[width=0.60\textwidth]{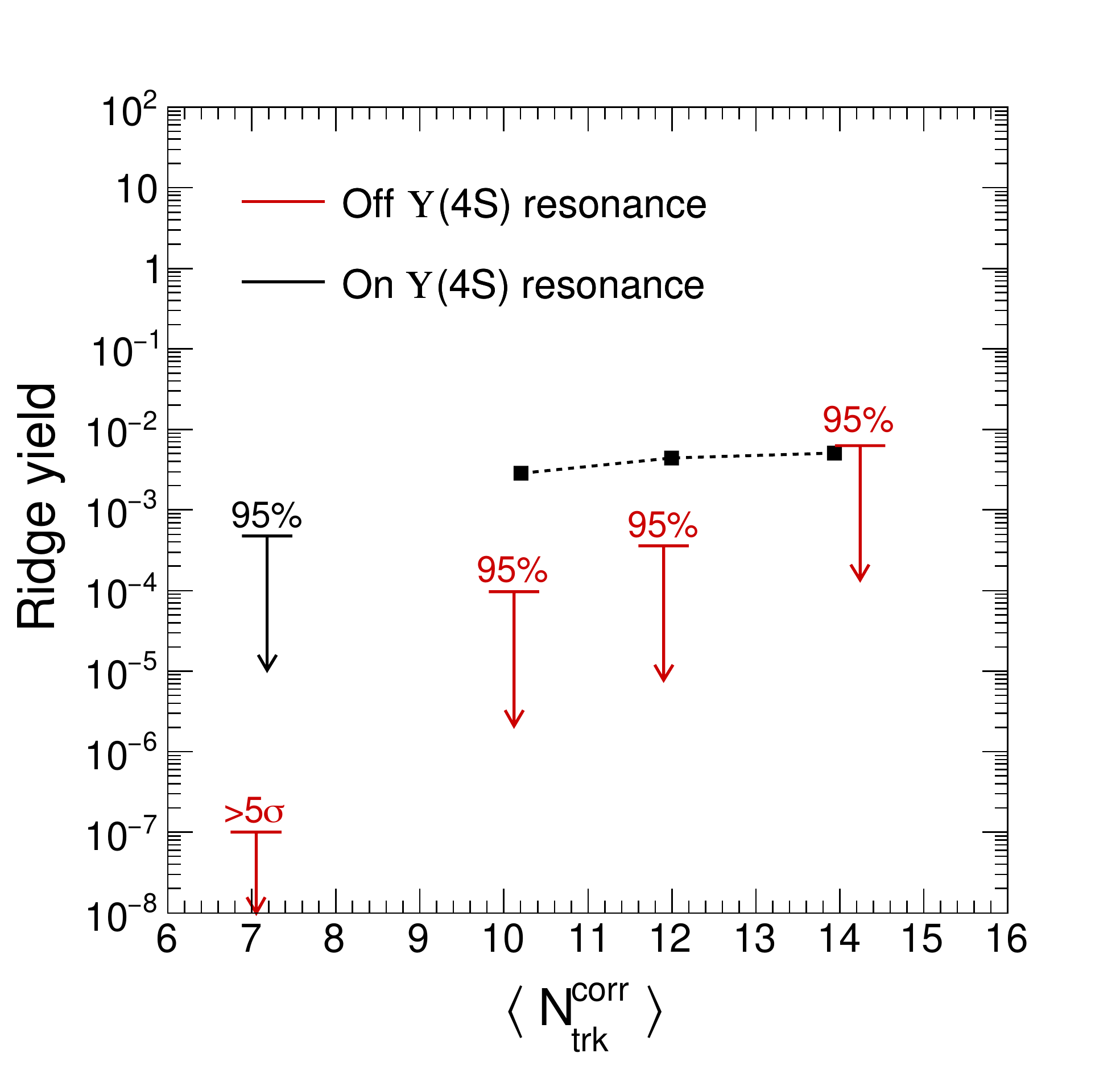}
\caption{
Central values (square markers for \onres) or upper limits of the ridge yield as a function of $\langle {\rm N}_{\rm trk}^{\rm corr}\rangle$ for the \conti (red) and \onres (black) datasets for the thrust axis analyses. 
The associated uncertainties with the measured ridge-yield central values are smaller than the symbols.
The label ``$>5\sigma$'' indicates a $5 \sigma$ confidence level upper limit.}
\label{fig:CLplots}
\end{figure}

For the measurement shows no obvious ridge-like structure, the ridge yield upper limit is reported using the bootstrap method~\cite{Efron:1979bst}.
The measured azimuthal differential associated yields are varied according to their statistical and systematic uncertainties and re-evaluated with the ZYAM method to obtain an alternative ridge yield.
The procedure is repeated two million times to form the bootstrapped ridge yield datasets using the three fit templates (see eq.~(\ref{eqn:ZYAMFuncs})), where the most conservative 95\% upper limits are reported. 
We report 95\% upper limits of the ridge yield or the confidence level of ridge-signal exclusion at $10^{-7}$ as a function of the averaged corrected multiplicity $\langle {\rm N}_{\rm trk}^{\rm corr}\rangle$.
With the large Belle data size, strong $5\sigma$ ridge-signal exclusions at $10^{-7}$ are set in the beam axis analysis in most multiplicity ranges for both the \conti and \onres data, except the highest multiplicity bin (${\rm N}_{\rm trk}^{\rm corr} \ge 14$) with the least statistics in the \conti sample, a $97\%$ confidence level of $10^{-7}$ exclusion is set.
The upper limit results of the thrust axis analysis are summarized in figure~\ref{fig:CLplots}. 
The \conti upper limits are less than $O(10^{-3})$ level for the majority of events.
The smallness of \onres long-range near-side enhancement, as pointed out in figure~\ref{fig:OnresThrust2PCProj}, is quantified at about $O(10^{-3})$ to $O(10^{-2})$. Upper limits are also provided considering the quantitative significance level of the measured central values.

\section{Studies with event generators}
\label{sec:GeneratorStudy}
\begin{figure}[ht]
\centering
	\begin{subfigure}[b]{\textwidth}
	\centering
		\includegraphics[width=0.95\textwidth]{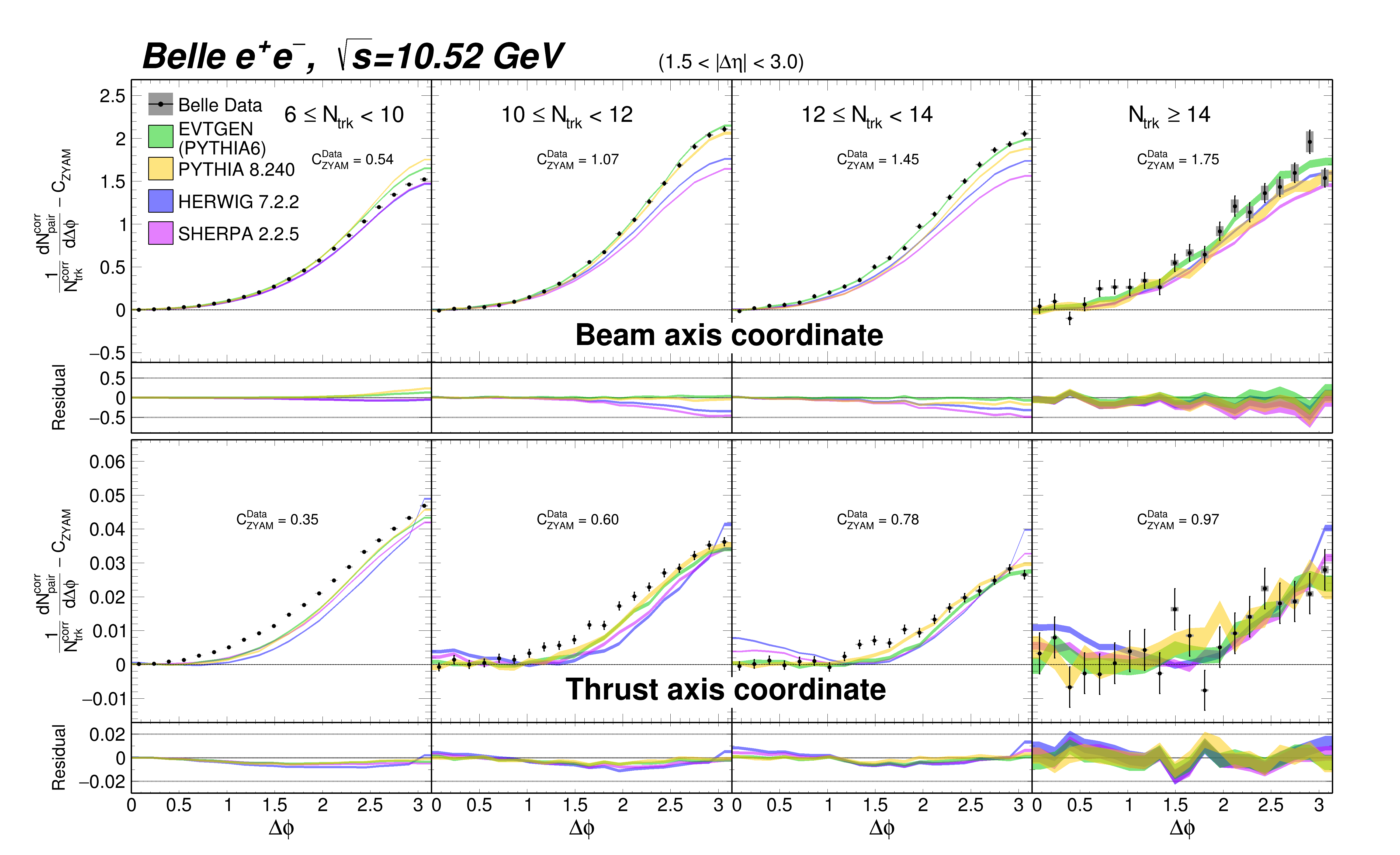}
		\caption{off-resonance}
		\label{fig:dNdphi_generator_conti}
	\end{subfigure}
    \begin{subfigure}[b]{\textwidth}
    \centering
		\includegraphics[width=0.95\textwidth]{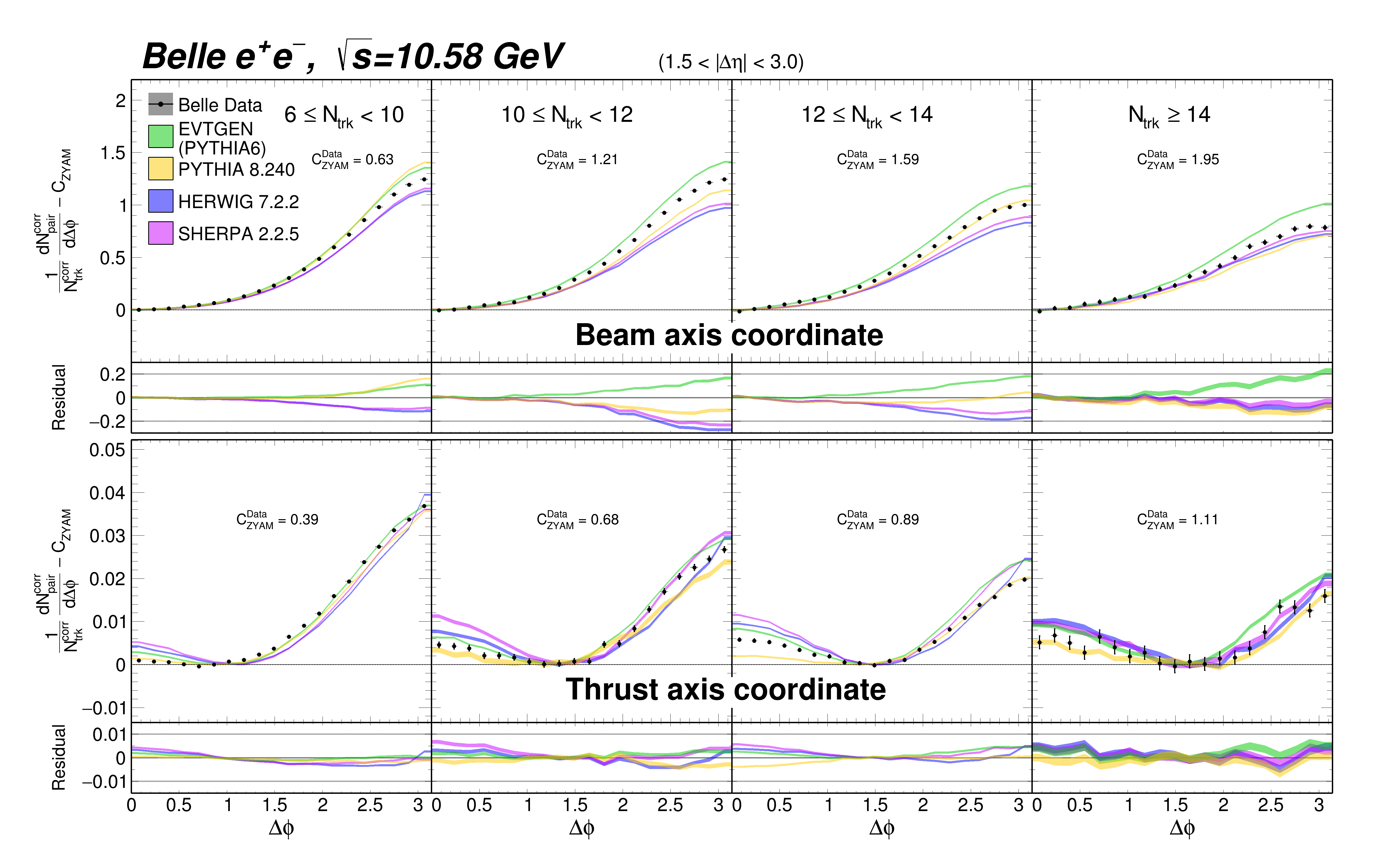}
        \caption{on-resonance}
		\label{fig:dNdphi_generator_onres}
    \end{subfigure}
\caption{Comparison of the ZYAM-subtracted $Y(\Delta \phi)$ in the range $1.5 \le |\Delta \eta| < 3.0$ for beam (top) and thrust (bottom) axis analyses as a function of the charged-particle multiplicity. 
The colored bands show simulation predictions from \textsc{pythia 6} (green), \textsc{pythia8} (yellow), \textsc{herwig} (blue) and \textsc{sherpa} (violet). Reconstruction effects on data (data points) are corrected, with error bars representing associated statistical uncertainties, and the gray boxes the systematic uncertainties.
For visual purposes, the minimal statistical uncertainties of the MC correlation colored bands are set to be $0.4\%$ of the plotting ranges, and residual panels have minimum thresholds of $3\%$ of the plotting ranges.}
\label{fig:dNdphi_generator}
\end{figure}

Complementary to ref.~\cite{Belle:2022fvl}, we provide the \conti and \onres results in full \ntrkoff analysis ranges comparing data with \textsc{pythia6}-based Belle MC, \textsc{pythia 8.240}~\cite{Sjostrand:2014zea}, \textsc{herwig 7.2.2}~\cite{Bellm:2015jjp} and \textsc{sherpa 2.2.5}~\cite{Gleisberg:2008ta} event generators for long-range azimuthal differential associated yields after performing the ZYAM method.
In this sub-section, we will call the \textsc{pythia6}-based Belle MC  by its version number \textsc{pythia6}, in comparison with other event generators. Figure~\ref{fig:dNdphi_generator} displays the results.

Generators (\textsc{pythia8}, \textsc{herwig} and \textsc{sherpa}) compared here are configured based on their default settings. More specifically, in \textsc{pythia8} event generation, the Monash 2013 tune~\cite{Skands:2014pea} ($e^+e^-$ collision's default tune) is used. The $e^+e^- \to q\bar{q} $ annihilation process is simulated with the full interference between photon and $Z$-boson propagators (option \texttt{WeakSingleBoson:ffbar2gmZ}), and the on-resonance production is generated by decaying the $\Upsilon(4S)$ particle evenly into charged and neutral $B$-meson pairs, where the mass and decay width of $\Upsilon(4S)$ are set to the corresponding world-average values~\cite{PDG}.

In the \textsc{herwig} event generation,  we require the order of electroweak coupling to be 2 (\texttt{OrderInAlphaEW}=2) and the order of strong coupling to be 0 (\texttt{OrderInAlphaS}=0) in the hard process simulation. For the $\Upsilon(4S)$ resonance production, the built-in matrix element for $e^+e^-$ colliding into a spin-$1$ vector meson (\texttt{MEee2VectorMeson}) is used.  
Decays of $B$ mesons are further done by the \textsc{evtgen/pythia8} interface.

The \textsc{sherpa} MC samples are generated with $\alpha_{\rm S} (M_Z) = 0.1188$, and a two-loop correction of the running of $\alpha_{\rm S}$ (\texttt{ORDER\_ALPHAS}=1) is specified. Full mass effects of charm and beauty quarks are considered. For the $\Upsilon(4S)$ resonance production, charged and neutral $B$-meson pairs are generated with the same probabilities.

In general, there is no significant indication of  multiplicity-dependent trends in the comparisons of data and MC long-range correlations, while minor inconsistencies across different multiplicity ranges could come from imperfections of the simulated multiplicity distributions.
As pointed out in ref.~\cite{Belle:2022fvl} using  high-multiplicity results, \textsc{pythia6} shows good agreements with data in the \conti results (figure~\ref{fig:dNdphi_generator_conti}). \textsc{herwig} and \textsc{sherpa} generators are also capable of describing the data correlations in the near-side region, but have discrepancies in the away-side correlation magnitudes. In the thrust-axis analysis, there are differences in data and MC correlation distribution shapes. 
MCs have flatter correlations in the near-side region, while their away-side peaks are steeper than that of data. 
Besides this general behavior, a notable feature is that \textsc{herwig} overshoots data as $\dphi < 1$ and $\dphi \approx \pi$; \textsc{sherpa} shows similar but smaller deviations.

Similarly, disagreements in the away-side correlation magnitude also appear in the \onres beam-axis results for \textsc{pythia8}, \textsc{herwig}, and \textsc{sherpa} generators in figure~\ref{fig:dNdphi_generator_onres}. But unlike the \conti case, there are over-predictions in the beam-axis away-side magnitude from \textsc{pythia6}.
There is mild near-side enhancement in the \onres thrust-axis correlations. This feature is also found in generator results, with \textsc{pythia6}, \textsc{herwig}, and \textsc{sherpa} overshooting the data points, while \textsc{pythia8} undershoots them.

\clearpage

\subsection{The thrust-axis origin-peak jet correlations with simulation}
\label{sec:ContiOriPeak}
\begin{figure}[ht]
\centering
\includegraphics[width=\textwidth,trim={0 1cm 0 0},clip]{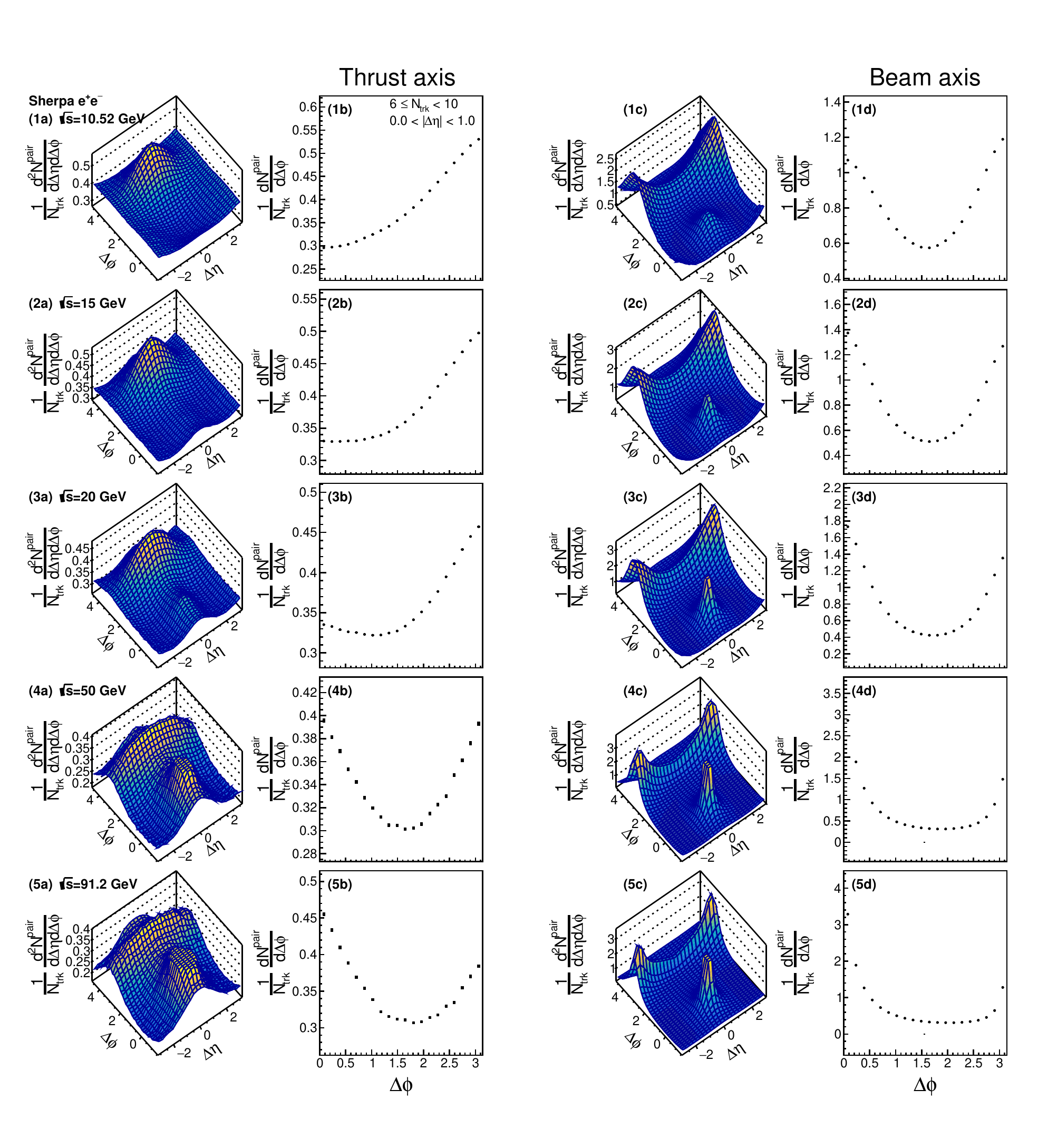}
\caption{Thrust-axis two-particle correlation functions and the short-range ($0 \le |\deta| < 1$) associated yield projections (left two columns) for \ee annihilation simulated by \textsc{sherpa} for different collision energies. The increasing trend of origin-peak correlations relative to the collision energy is observed regardless of the event multiplicity, and $\ntrkoff \in [6,10)$ is shown as representative here. 
For comparison, beam-axis two-particle correlations under each energy benchmark are shown in the right two columns.}
\label{fig:CollisionEnergy}
\end{figure}

As pointed out in section~\ref{sec:Rst}, the result with Belle data in the thrust axis analysis lacks a significant origin-peak correlation.
We study this behavior with the \textsc{Sherpa} event generator, simulating \ee annihilation events with collision energies ranging from 10.52 \gev up to the $Z$-pole energy.
The corresponding results are shown in figure~\ref{fig:CollisionEnergy}.
Initial state radiation is turned off in this simulation. For different collision energies, radiation photons' momentum kicks have different levels of recoil effects on $q\bar{q}$ fragmentation events, which is an additional unwanted complex factor for the correlation trend study as a function of fragmenting quarks' energy. 
However, a consistent evolving trend of the magnitude of the origin-peak correlation is seen with increasing collision energy for every multiplicity interval.
Under the picture in the thrust axis analysis, the main contribution to the origin-peak correlation is contributed from jets fragmenting in the transverse direction of the event thrust axis.
Increasing the collision energy can also make these jets more energetic.

\subsection{The thrust-axis long-range near-side enhancement of on-resonance data}
\label{sec:OnresRidge}
We observe an enhancement of long-range near-side correlations with the \onres data, as reported in section~\ref{sec:Rst}. 
MC simulation also reproduces a similar order of long-range near-side excess; hence, we study this effect with the on-resonance Belle MC sample to further understand the impact on two-particle correlation functions arising from $B$-decay events. 

\begin{figure}
\centering
\includegraphics[width=\textwidth,trim={0 1cm 0 0},clip]{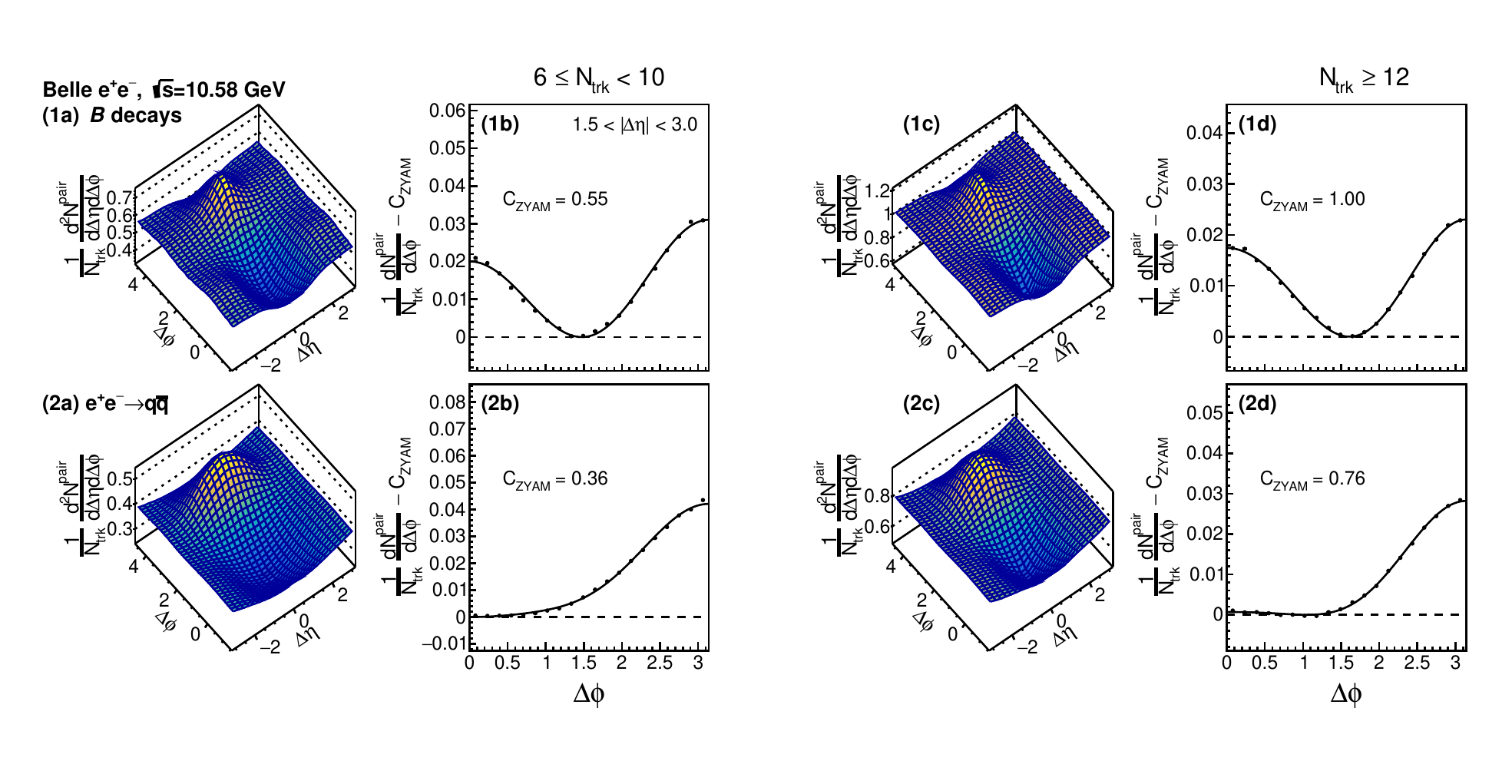}
\caption{Two-particle correlation functions and long-range ($1.5 \le |\deta| < 3.0$) ZYAM-subtracted azimuthal differential associated yields for separated MC sources with the Belle MC. Panels (1a)-(1d) are results with $B$-meson decays and panels (2a)-(2d) are those from $q\bar{q}$ fragmentation, in which (a) and (b) show the lower multiplicity results with $6 \le N_{\rm trk} < 10$; (c) and (d) show the higher multiplicity ones with $N_{\rm trk} \ge 12$. The long-range near-side enhancements are seen specifically in the $B$ decays MC sample in both the low- and high-multiplicity cases.}
\label{fig:RecoOnresThrust2PC_BBQQ}
\end{figure}

We show for consistency the two-particle correlations and  ZYAM-subtracted associated yields with separated \B-decay and \qq fragmenting processes in figure~\ref{fig:RecoOnresThrust2PC_BBQQ}, based on the truth level MC.
As per expectation, the excess is mainly originated from \B-decay.
However, as mentioned in section~\ref{sec:Rst}, the enhanced structures seen for these $B$-decay events (or for that matter the on-resonance sample) have an intrinsic difference in shape from those reported with heavy ions collisions.
Recall for the latter, the typical ridge structure is an elongated feature spanning a wide pseudorapidity range, but the enhancement seen in the $B$-decay event sample are lumps only located at the long range regions. 
On the other hand, an anisotropy in azimuthal angle for charged particle distributions is also not anticipated in the $B$ decay system, which is believed to be well described by the decay relations and energy-momentum conservation. 
To understand the long-range near-side enhancement in the $B$-decay events under the thrust axis analysis, a decomposition study of the correlation contribution among two $B$ mesons is performed.
In the following, we examine the correlation function with track pairs selected specifically from:
\begin{enumerate}
\item Single $B$ meson decay products:\\
The associated particle and the trigger particle should be descendent from the same $B$ meson.
\item Different $B$ meson decay products:\\
The associated particle and the trigger particle should be descendant from opposite $B$ mesons.
\end{enumerate}

\begin{figure}[ht]
\centering
\includegraphics[width=0.5\textwidth,trim={0 1cm 0 0},clip]{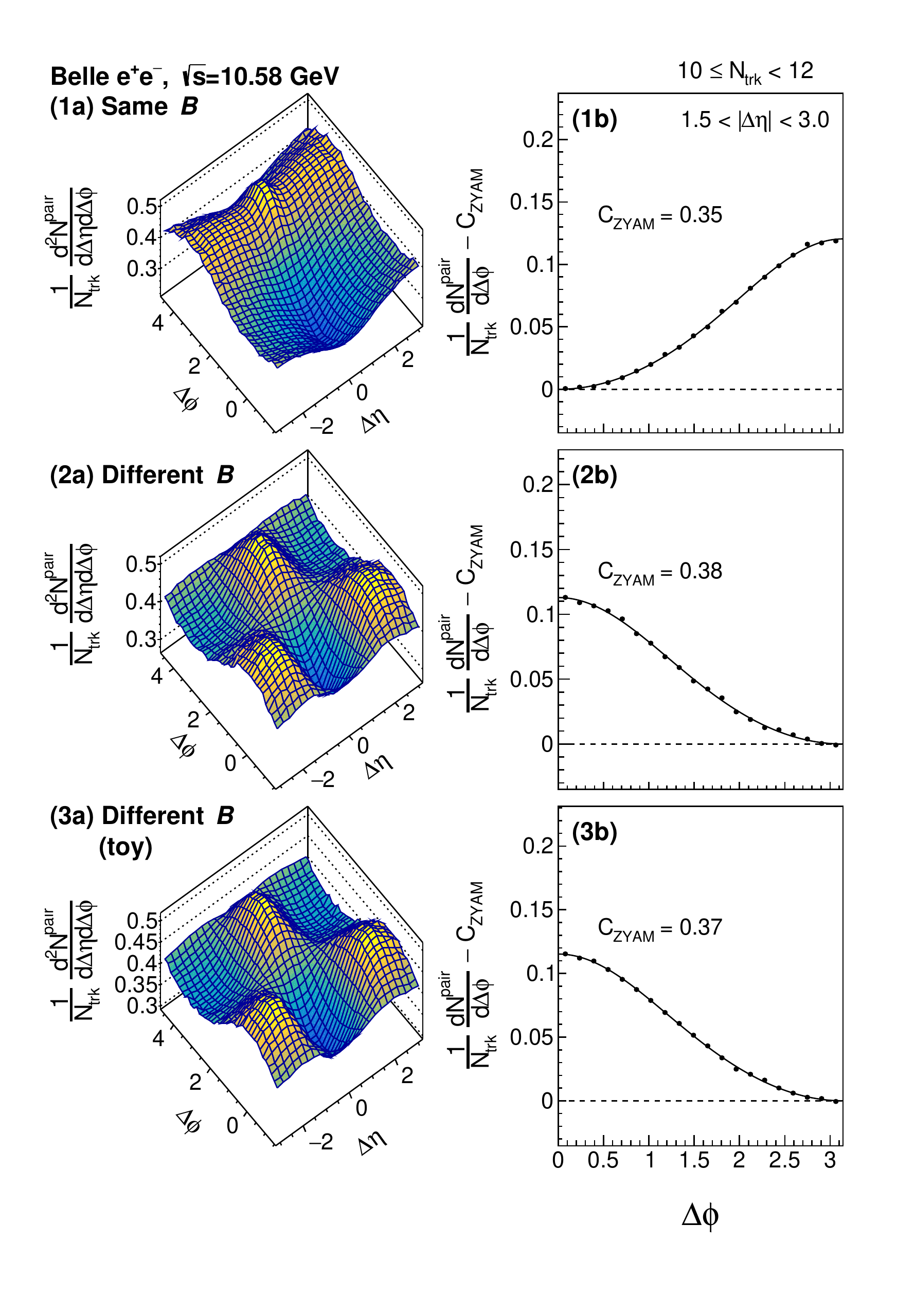}
\caption{Two-particle correlation functions and long-range ($1.5 \le |\deta| < 3.0$) ZYAM-subtracted azimuthal differential associated yields for tracks (1) from the same or (2) across different $B$ mesons for every $\Upsilon(4S) \to B\bar{B}$ event with the Belle MC. Panel (3) shows the pairing yields counted from different $B$ mesons from the toy sample mixed from uncorrelated events.}
\label{fig:RecoOnresThrust2PC_SameBOppositeB}
\end{figure}

\begin{figure}[ht]
\centering
    \includegraphics[width=0.6\textwidth, trim={0 1.5em 0 0},clip]{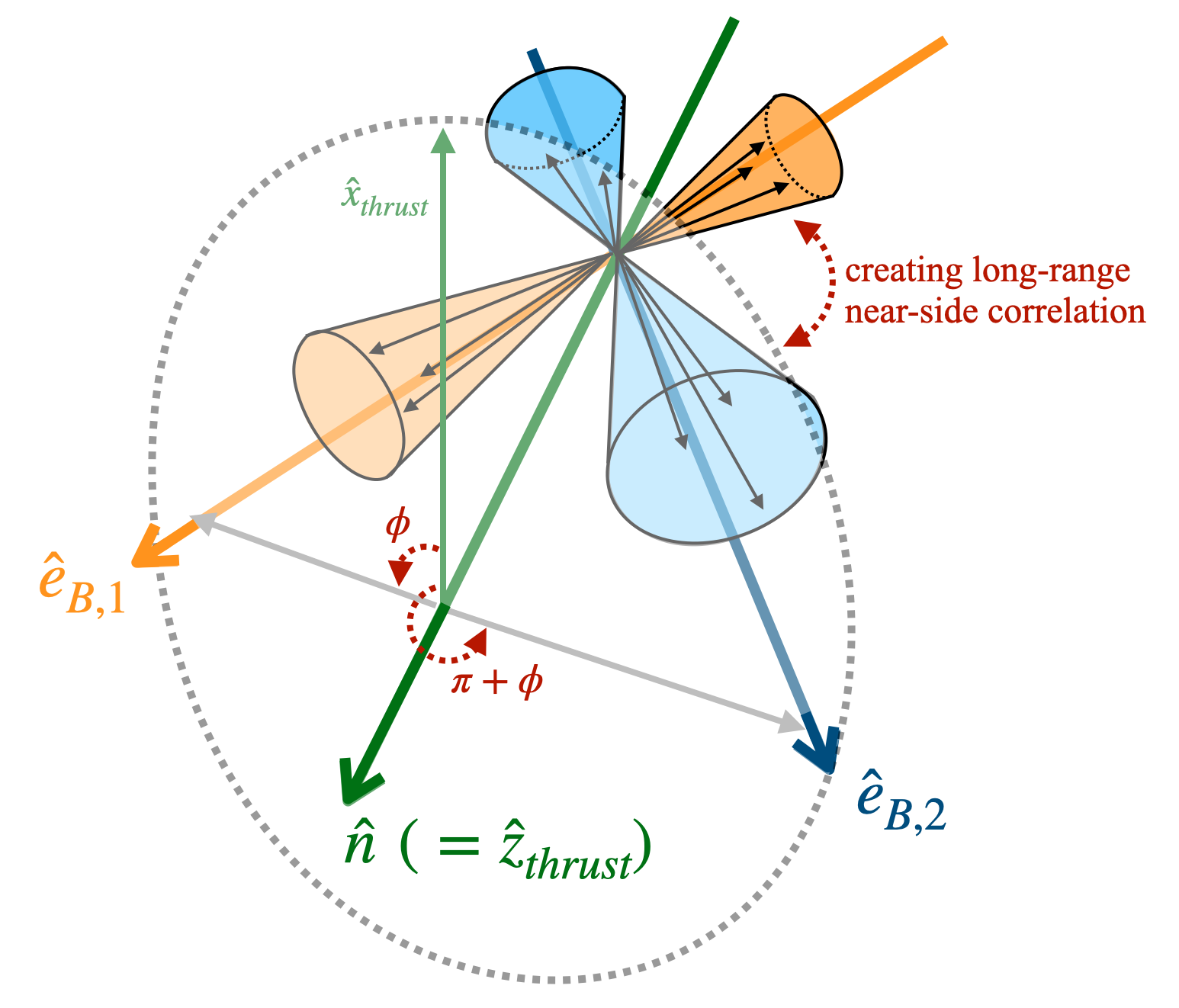}
\caption{
A schematic diagram indicating how a special arrangement of two \B decay system can have a long-range near-side correlation.
The orange cones represent the decay products from one of the $B$ mesons, and the blue cones are from the other $B$ decay. 
Two \B decay directions are expressed with notations $\hat{e}_{B,1}$, $\hat{e}_{B,2}$.
The event thrust direction $\hat{n}$  is prone to lie on the plane spanned by $\hat{e}_{B,1}$, $\hat{e}_{B,2}$, hence dividing the decay products into two parts, aggregating at azimuthal coordinates around $\phi$ or $\pi+\phi$, respectively. 
In this configuration, the decay products from the opposite $B$ mesons (the orange and blue cones) easily contribute to the long-range near-side correlation.
}
\label{fig:Onres_SameBOppositeB_explain}
\end{figure}

In figure~\ref{fig:RecoOnresThrust2PC_SameBOppositeB}, distinct patterns in the two-particle correlation functions and the long-range ZYAM-subtracted yields are observed under these two circumstances. 
The sizable enhancement in the long-range near-side region is seen exclusively in the correlation function calculated by track pairs from different $B$ mesons. 
A complementary exercise with a toy sample is checked in figure~\ref{fig:RecoOnresThrust2PC_SameBOppositeB}(3a)-\ref{fig:RecoOnresThrust2PC_SameBOppositeB}(3b). 
The toy sample is blended with two $B$ mesons picked up from uncorrelated events, and the event thrust axis is recalculated with the new mixture of constituents.
The calculation of the correlation function with track pairs from different $B$ mesons’ decay products using this toy sample dataset guarantees that there is no special physical correlation across these two $B$ meson's decay products. The only correlation in the blended toy sample is the momentum balance for the individual $B$ mesons' decays.
Therefore, the result concerns only the decay topology and the alignment of the thrust axis with respect to two $B$ decay systems.
The blended toy sample replicates the correlation function structure seen in crossing contribution in physical $B$-decay events (figure~\ref{fig:RecoOnresThrust2PC_SameBOppositeB}(2a)-\ref{fig:RecoOnresThrust2PC_SameBOppositeB}(2b)).
We provide a schematic explanation of enhanced near-side correlations under such $B$-meson-pair-decay events in figure~\ref{fig:Onres_SameBOppositeB_explain}.
The derived event thrust axis for the two-$B$-decay event has a large probability to lie in the plane formed by two $B$ mesons' main decay directions. 
The special circumstances further makes most of the constituents sit on either $\phi$ or $\pi+\phi$, yielding aggregated correlation at $\Delta \phi \approx 0$ or $\Delta \phi \approx \pi$.
It is concluded that the sculpted ridge signal in the $B$-decay events (or the on-resonance sample) is a consequence of the special event topology and thrust axis alignment in two $B$ meson decay systems.

\section{Conclusions}
\label{sec:conclusions}
This paper details the methodology and results of the two-particle correlations analysis in thrust axis coordinates. The method has been first utilized in the measurement with ALEPH data~\cite{Badea:2019vey} and subsequently adopted in the Belle measurement~\cite{Belle:2022fvl}.
To express particle correlations for a randomly oriented event topology, such as arising in the \ee collision system, we suggest that this thrust axis analysis is the more appropriate coordinate system compared to conventional fixed beam axis coordinates.
The thrust axis provides an estimate of the $\ee \to \qq$ direction in the hard process.
Constructed upon the thrust-axis, two-particle correlations for $\ee \to \qq$ events are an observable sensitive to higher-order soft emission transverse to the leading dijet axis.

In a broader view, we report full results of two-particle correlations in both the beam and thrust axis coordinate systems performed on datasets of $e^+e^-$ to hadronic final states at $\sqrt{s}= 10.52$~GeV and \sqrts=10.58~GeV. This amplifies the preceding Belle publication~\cite{Belle:2022fvl}, which focuses on the pure $\ee \to \qq$ fragmentation dataset at $\sqrt{s}=10.52$~GeV.
As already pointed out in that Letter, the thrust-axis correlation function distribution is different from measurements in hadron collisions and high-energy $e^+e^-$ collisions in its shape, with the significant near-side peak structure, interpreted as intra-jet correlation, not being seen.
In this paper, we research qualitatively the correlation structure taking advantage of simulations.
By studying with the \textsc{sherpa} event generator, we characterize the evolution of the near-side-peak correlation magnitude from low-energy collisions towards the high collision energies.

The two-particle correlation measurement with the $\Upsilon(4S)$ on-resonance dataset is presented for the first time. A low-scaled long-range near-side enhancement is observed in data for the thrust-axis two-particle correlations. However, this enhancement is different from the more familiar elongated ridge structure over a broad \deta range as reported in $pp$ collisions and heavy ion experiments.
Monte Carlo simulations are able to qualitatively replicate these new features of the on-resonance correlation data. It is concluded that such a special topological arrangement within a decay system can give rise to such correlations.

\section*{Acknowledgements}
This work, based on data collected using the Belle detector, which was
operated until June 2010, was supported by 
the Ministry of Education, Culture, Sports, Science, and
Technology (MEXT) of Japan, the Japan Society for the 
Promotion of Science (JSPS), and the Tau-Lepton Physics 
Research Center of Nagoya University; 
the Australian Research Council including grants
DP180102629, 
DP170102389, 
DP170102204, 
DE220100462, 
DP150103061, 
FT130100303; 
Austrian Federal Ministry of Education, Science and Research (FWF) and
FWF Austrian Science Fund No.~P~31361-N36;
the National Natural Science Foundation of China under Contracts
No.~11675166,  
No.~11705209;  
No.~11975076;  
No.~12135005;  
No.~12175041;  
No.~12161141008; 
Key Research Program of Frontier Sciences, Chinese Academy of Sciences (CAS), Grant No.~QYZDJ-SSW-SLH011; 
the Ministry of Education, Youth and Sports of the Czech
Republic under Contract No.~LTT17020;
the Czech Science Foundation Grant No. 22-18469S;
Horizon 2020 ERC Advanced Grant No.~884719 and ERC Starting Grant No.~947006 ``InterLeptons'' (European Union);
the Carl Zeiss Foundation, the Deutsche Forschungsgemeinschaft, the
Excellence Cluster Universe, and the VolkswagenStiftung;
the Department of Atomic Energy (Project Identification No. RTI 4002) and the Department of Science and Technology of India; 
the Istituto Nazionale di Fisica Nucleare of Italy; 
National Research Foundation (NRF) of Korea Grant
Nos.~2016R1\-D1A1B\-02012900, 2018R1\-A2B\-3003643,
2018R1\-A6A1A\-06024970, RS\-2022\-00197659,
2019R1\-I1A3A\-01058933, 2021R1\-A6A1A\-03043957,
2021R1\-F1A\-1060423, 2021R1\-F1A\-1064008, 2022R1\-A2C\-1003993;
Radiation Science Research Institute, Foreign Large-size Research Facility Application Supporting project, the Global Science Experimental Data Hub Center of the Korea Institute of Science and Technology Information and KREONET/GLORIAD;
the Polish Ministry of Science and Higher Education and 
the National Science Center;
the Ministry of Science and Higher Education of the Russian Federation, Agreement 14.W03.31.0026, 
and the HSE University Basic Research Program, Moscow; 
University of Tabuk research grants
S-1440-0321, S-0256-1438, and S-0280-1439 (Saudi Arabia);
the Slovenian Research Agency Grant Nos. J1-9124 and P1-0135;
Ikerbasque, Basque Foundation for Science, Spain;
the Swiss National Science Foundation; 
the Ministry of Education and the Ministry of Science and Technology of Taiwan;
and the United States Department of Energy and the National Science Foundation.
These acknowledgements are not to be interpreted as an endorsement of any
statement made by any of our institutes, funding agencies, governments, or
their representatives.
We thank the KEKB group for the excellent operation of the
accelerator; the KEK cryogenics group for the efficient
operation of the solenoid; and the KEK computer group and the Pacific Northwest National
Laboratory (PNNL) Environmental Molecular Sciences Laboratory (EMSL)
computing group for strong computing support; and the National
Institute of Informatics, and Science Information NETwork 6 (SINET6) for
valuable network support.

\clearpage
\bibliographystyle{JHEP}
\bibliography{TPC}
\end{document}